\documentclass[journal,twoside]{IEEEtran}
\usepackage{times,epsfig}
\usepackage{cite}
\usepackage{graphicx}
\usepackage[cmex10]{amsmath}
\usepackage{url}
\usepackage{multirow}
\usepackage{epstopdf}
\usepackage{algorithm}
\usepackage{algpseudocode}
\usepackage{amsmath}

\usepackage{colortbl}
\usepackage{algorithmicx}
\usepackage{pgfplots}
\pgfplotsset{compat=1.5}
\usepgfplotslibrary{groupplots}
\usepackage{hyperref}
\pgfplotsset{compat=1.5}
\usepgfplotslibrary{groupplots}
\usepackage{booktabs}
\usepackage{threeparttable}
\usepackage{tcolorbox}
\usepackage{tabularx}
\usepackage{xcolor}
\usepackage{graphicx}
\usepackage{float}
\usepackage{bm}
\usepackage[caption=false,font=footnotesize,labelfont=rm,textfont=rm]{subfig}

\begin{document}
\title{USTC-TD: A Test Dataset and Benchmark for Image and Video Coding in 2020s}
\author{
Zhuoyuan Li, 
Junqi Liao,
Chuanbo Tang,
Haotian Zhang,
Yuqi Li,
Yifan Bian,
Xihua Sheng,
Xinmin Feng,\\
Yao Li,
Changsheng Gao,
Li Li, \IEEEmembership{Member, IEEE,}
Dong Liu, \IEEEmembership{Senior Member, IEEE,}
and Feng Wu, \IEEEmembership{Fellow, IEEE} \vspace{-2em}

\thanks{Date of current version \today. \textit{(Z. Li and J. Liao contributed equally to this work.)} \textit{(Corresponding authors: Dong Liu and Feng Wu.)}

The authors are with the MOE Key Laboratory of Brain-Inspired Intelligent Perception and Cognition, University of Science and Technology of China, Hefei 230027, China (e-mail: \{zhuoyuanli, liaojq, cbtang, zhanghaotian, lyq010303, esakak, xhsheng, xmfeng2000, mrliyao\}@mail.ustc.edu.cn; \{changshenggao, lil1, dongeliu, fengwu\}@ustc.edu.cn).}}
\markboth{Under Review}
{Li \MakeLowercase{\textit{et al.}}: USTC-TD: A Test Dataset and Benchmark for Image and Video Coding in 2020s}

\maketitle

\begin{abstract}
Image/video coding has been a remarkable research area for both academia and industry for many years. Testing datasets, especially high-quality image/video datasets are desirable for the justified evaluation of coding-related research, practical applications, and standardization activities. We put forward a test dataset namely USTC-TD, which has been successfully adopted in the practical end-to-end image/video coding challenge of the \textit{IEEE International Conference on Visual Communications and lmage Processing (VCIP)} in 2022 and 2023. USTC-TD contains 40 images at 4K spatial resolution and 10 video sequences at 1080p spatial resolution, featuring various content due to the diverse environmental factors (\textit{e.g.} scene type, texture, motion, view) and the designed imaging factors (\textit{e.g.} illumination, lens, shadow). We quantitatively evaluate USTC-TD on different image/video features (spatial, temporal, color, lightness), and compare it with the previous image/video test datasets, which verifies its excellent compensation for the shortcomings of existing datasets. We also evaluate both classic standardized and recently learned image/video coding schemes on USTC-TD using objective quality metrics (PSNR, MS-SSIM, VMAF) and subjective quality metric (MOS), providing an extensive benchmark for these evaluated schemes. Based on the characteristics and specific design of the proposed test dataset, we analyze the benchmark performance and shed light on the future research and development of image/video coding. All the data are released online: \url{https://esakak.github.io/USTC-TD}\,.
\end{abstract}

\IEEEpeerreviewmaketitle

\vspace{-0.5em}
\section{Introduction}
\vspace{-0.1em}
Nowadays, with the dramatic growth of data traffic over the internet and the emergent application of versatile image/video formats such as 2K, 4K, high dynamic range, and wide color gamut, there is a pressing demand for storage and transmission. To address this challenge, in recent decades, image/video compression is employed to reduce the amount of data significantly, and several video coding standards have been developed, such as High Efficiency Video Coding (H.265/HEVC)\cite{sullivan2012overview}, Versatile Video Coding (H.266/VVC)\cite{bross2021overview}, Audio Video Standard (AVS1, AVS2, AVS3)\cite{ma2022evolution}, AOMedia Video 1 and 2  (AV1\cite{han2021technical}, AV2\cite{zhao2021study}).

End-to-end image/video compression has been a research focus on visual data compression for both academia and industry for over six years\cite{balle2018variational,minnen2018joint,cheng2020learned,ma2020end,he2022elic,jiang2023mlic,lu2019dvc, hu2021fvc, lu2020content, lu2022learning, ho2022canf, chen2023b, chen2023canf, li2021deep, sheng2022temporal, li2022hybrid, tang2024offline, sheng2024spatial, bian2024lssvc, sheng2024vnvc, li2023neural, li2024neural,chen2021nerv,chen2023hnerv,kwan2024hinerv,kwan2024nvrc}. A number of technologies have been developed, such as expressive auto-encoder neural networks, precise probability estimation neural networks, and conditional end-to-end coding frameworks, and so on. Until recently, the performances of both end-to-end image and video compression schemes have surpassed that of the advanced H.266/VVC under certain test conditions \cite{he2022elic,jiang2023mlic, li2023neural,li2024neural}. 

For the evaluation of these image/video compression schemes in the practical application and standardization, they are usually benchmarked with objective and subjective quality metrics to evaluate their rate-distortion (RD) performance and a trade-off between coding efficiency and reconstructed quality. To sufficiently consider the effectiveness of these quality assessments, the test results of representative image/video test datasets are the key to reflecting the practicability and generalization of the researcher's scheme.

In this paper, a new image/video dataset, named USTC-TD, is proposed for testing and evaluating the practical image/video coding algorithms. USTC-TD contains 40 images and 10 video sequences with a wide content coverage. For the image dataset, each image is captured with a high spatial resolution (4K) and converted into RGB, YUV444/420 color space, and PNG/YUV file format. For the video dataset, each video sequence consists of 96/300 frames, and each frame is captured at 30 frames per second (fps) with 1080p spatial resolution and converted into RGB, YUV444/420 color space, and PNG/YUV file format. For the construction of image and video datasets, the data is collected with the specific design of different content factors (environmental/imaging-related factors), which aims to cover as close as possible to the real-world coding transmission scenes. Compared with the common test image/video datasets \cite{Kodak,asuni2014testimages,clic,mercat2020uvg,wang2016mcl,bossen2013common, boyce2018jvet}, we use different quantitative criteria to comprehensively evaluate the diversity of the proposed USTC-TD  from the perspective of spatial, temporal, colorfulness, and lightness information, demonstrating its excellent coverage and effectively compensating for the shortcomings of existing datasets.

\begin{table*}
	\renewcommand\arraystretch{1.2}
	\centering
	\scriptsize
	\vspace{-2em}
	\caption{Common Test Datasets of Image Compression}
	\vspace{-0.8em}
	\label{tab:LICB}
	\setlength{\tabcolsep}{3mm}
	{
		\begin{tabular}{c|c|c|c|c|c|c}
			\hline
			\textbf{Dataset} & \textbf{Resolution}                  & \multicolumn{1}{l|}{\textbf{Number}} & \textbf{Color Space} & \textbf{Bit Depth} & \textbf{Setting}            & \textbf{Characteristic}                                                                                        \\ \hline
			\textit{Kodak}\cite{Kodak}   & 768×512                              & 24                                   & RGB                  & 24                 & - & \multirow{3}{*}{\begin{tabular}[c]{@{}c@{}}Rich Texture\\ Various Scenery\\ Appropriate Exposure\end{tabular}} \\
			\textit{Tecnick}\cite{asuni2014testimages} & 1200×1200                            & 100                                  & RGB                  & 24                 & Sampling                    &                                                                                                                \\
			\textit{CLIC}\cite{clic}    & \multicolumn{1}{l|}{1189×1789 (AVG)} & 41                                   & RGB                  & 24                 & Valid-Professional          &                                                                                                                \\ \hline
		\end{tabular}
	}
\end{table*}

\begin{table*}
	\renewcommand\arraystretch{1.1}
	\centering
	\scriptsize
	\vspace{-1em}
	\caption{Common Test Datasets of Video Compression}
	\vspace{-0.8em}
	\label{tab:LVCB}
	\setlength{\tabcolsep}{4mm}
	{
		\begin{tabular}{c|c|c|c|c|c}
			\hline
			\textbf{Dataset}  & \textbf{Resolution} & \multicolumn{1}{l|}{\textbf{Number}} & \textbf{FPS} & \textbf{Length}                     & \textbf{Characteristic}                                                                                                 \\ \hline
			\textit{\textbf{UVG}}\cite{mercat2020uvg}      & 3840×2160           & 16                                   & 50/120       & 5s-12s                              & \multirow{3}{*}{\begin{tabular}[c]{@{}c@{}}Fast/Slow Motion\\ Diverse Video Scenes\\ Appropriate Exposure\end{tabular}} \\
			\textit{\textbf{MCL-JCV}}\cite{wang2016mcl}  & 1920×1080           & 30                                   & -            & 5s                                  &                                                                                                                         \\
			\textit{\textbf{HEVC CTC}}\cite{bossen2013common}, \textit{\textbf{VVC CTC}}\cite{boyce2018jvet} & 240P-4K             & 41                                   & 24-60        & \multicolumn{1}{l|}{150-600 Frames} &                                                                                                                         \\ \hline
		\end{tabular}
	}
 \vspace{-1.5em}
\end{table*}

In addition, we establish baselines and evaluate the classic standardized compression schemes \cite{sullivan2012overview,bross2021overview,ma2022evolution,han2021technical,zhao2021study} and recently learned image/video compression schemes \cite{balle2018variational,minnen2018joint,cheng2020learned,ma2020end,he2022elic,jiang2023mlic,lu2019dvc,ho2022canf, li2021deep, sheng2022temporal,li2022hybrid, tang2024offline,sheng2024spatial,sheng2024vnvc,li2023neural,li2024neural,chen2021nerv,chen2023hnerv,kwan2024hinerv,kwan2024nvrc} under objective quality metrics (\textit{PSNR}, \textit{MS-SSIM}\cite{wang2003multiscale}, \textit{VMAF}\,\footnote{\vspace{-0.4em}Available online at \,\url{https://github.com/Netflix/vmaf}.}\cite{liu2013visual}) and subjective quality metric (\textit{MOS}) \cite{rec2006p, international1996methods, itu2017vocabulary, series2012methodology}), and then benchmark and analyze their performance on the proposed dataset to shed light on future research and development of image/video coding. The benchmark data and test scripts are available online with the proposed dataset and released on the open-sourced website for researchers to reproduce conveniently. 

We hope the proposed test datasets allow researchers to make more well-informed decisions under efficient evaluation, and guide the innovation and improvement of future schemes and experiments. In summary, our contributions are as follows:
\begin{itemize}
\item We build a new image/video compression test dataset (named USTC-TD), which focuses on the diversity of various content factors.

\item We conduct a comprehensive evaluation of the proposed dataset by using different quantitative criteria,  demonstrating the excellent compensation of USTC-TD for existing image/video datasets.

\item We conduct a comprehensive evaluation of the advanced image/video compression schemes on the proposed dataset, and establish an extended baseline for the evaluative image/video coding schemes benchmarked on USTC-TD.

\item Taking a close look at USTC-TD, we analyze the benchmarked performance and shed light on the future research and development of image/video coding.
\end{itemize}

The remainder of the paper is structured as follows. Section II mentions the background of previous compression-related test datasets. Section III summarises the data collection process of the proposed dataset. Section IV introduces the construction of the image and video dataset of USTC-TD, and discusses the characteristics and utilization of the proposed dataset. Section V presents the experimental configuration and the evaluation of the advanced compression schemes on USTC-TD, and further analyzes their performance, limitation, and inspiration. Section VI concludes the paper and presents some suggestions for future work.

\vspace{-1em}
\section{Background}
\vspace{-0.2em}
In the past twenty years, with the rapid development of multimedia data over the advanced exhibition devices, resolutions, frame rates, dynamic range and viewpoints, the transmission/storage quantity of multimedia data is progressively accompanied by dramatic increases in the requirement of users. As the powerful multimedia data transmission/storage tool, lossy/lossless image/video compression has become the primary driver for reducing the internet bandwidth and storage. For the standardization activities and research of compression-related systems, image/video test (evaluation) dataset is a critical component for optimizing the performance and reflecting the practicability and generalization of different compression schemes. Here we review the image/video test dataset commonly used by standards and researchers in the past, and summarize their characteristics.

{\bf Image Compression Test Dataset.} For the evaluation of previous image compression schemes, \textit{Kodak}\,\footnote{Available online at \,\url{https://r0k.us/graphics/kodak/}.}\cite{Kodak}, \textit{Tecnick}\cite{asuni2014testimages} (sampling setting), \textit{CLIC} (professional setting) \cite{clic} are commonly used, the setting is mentioned in Table~\ref{tab:LICB}\,, and the introduction is summarised below: 

\begin{itemize}
\item \textbf{\textit{Kodak}}\cite{Kodak} is a commonly used true color set of images released for various testing purposes and benchmarks, it contains 24 images with RGB format. The images are all photographic type and continuous tone. Many sites use them as a standard test suite for compression testing.

\item \textbf{\textit{Tecnick}}\cite{asuni2014testimages} is a huge collection of sample images designed for quality assessment of different kinds of displays and image processing techniques. The sampling setting is widely used on testing resampling algorithms.

\item \textbf{\textit{CLIC}}\cite{clic} is 
a high-quality image set collected from \textit{Unsplash}\,\footnote{Available online at \,\url{https://unsplash.com/}.}, and contains images of similar quality from potentially different sources. It has been successfully applied in the workshop and challenge on learned image compression (\textit{CLIC}) of \textit{IEEE/CVF Computer Vision and Pattern Recognition Conference} (\textit{CVPR}) and \textit{Data Compression Conference} (\textit{DCC}).
\end{itemize}

In these image datasets, characteristics mainly focus on the different resolutions with more scene types, most of the test images are captured by high-definition lens in specific scenes. These datasets aim to evaluate the basic ability of image compression algorithms to remove intra-frame redundancy for different scenarios, but the limited diversity of content factors makes it difficult to evaluate the robustness of algorithms.

{\bf Video Compression Test Dataset.} For the evaluation of previous video compression schemes, \textit{UVG}\cite{mercat2020uvg}, \textit{MCL-JCV}\cite{wang2016mcl}, 
\textit{HEVC Common Test Conditions} (\textit{CTC})\cite{bossen2013common}, \textit{VVC CTC} \cite{boyce2018jvet} are commonly used, the settings are mentioned in the Table~\ref{tab:LVCB}\,, and the introduction is summarised below: 

\begin{itemize}
\item \textbf{\textit{UVG}}\cite{mercat2020uvg} contains 16 test video sequences. They are captured with Sony F65 video camera in 16-bit F65RAW-HFR format and converted to YUV420 videos by \textit{ffmpeg} tool\,\footnote{Available online at \,\url{https://ffmpeg.org/}.}. It is widely used in the evaluation of advanced learned video compression methods\cite{lu2019dvc,ho2022canf, li2021deep,sheng2022temporal,li2022hybrid,tang2024offline,sheng2024spatial,sheng2024vnvc,li2023neural,li2024neural,chen2021nerv,chen2023hnerv,kwan2024hinerv,kwan2024nvrc}.

\item \textbf{\textit{MCL-JCV}}\cite{wang2016mcl} is a compressed video quality assessment dataset based on the just noticeable difference (JND) model. All provided video sequences are available to the public with measured raw JND data for each test subject and allow users to do their own processing. 

\item \textbf{\textit{HEVC CTC}, and \textit{VVC CTC}}\cite{bossen2013common, boyce2018jvet} define the common test conditions and test sequences for the standardization activities of H.265/HEVC \cite{sullivan2012overview} and H.266/VVC \cite{bross2021overview}, and protect the core experiments in a well-defined rule. It promotes the upgrading of many technologies in standardization, and has been widely used in compression-related systems.
\end{itemize}

In these video test datasets, the characteristics mainly focus on the various video contents, including simple/complex motion and poor/high capture quality. Most of the test videos can only evaluate the basic ability of video compression-related algorithms to remove inter-frame redundancy for different scenarios under different video coding configurations, like motion estimation (ME), motion compensation (MC), and rate allocation/control (RC) technologies in low-delay (LD) and random access (RA) configurations, but these video contents with the limited types of temporal correlation make it difficult to evaluate the robustness of temporal property-related algorithms in video-based compression applications.

\begin{table}
	\renewcommand\arraystretch{1.3}
	\centering
	\scriptsize
	\caption{Camera Captured Parameters of USTC-TD 2022}
	\vspace{-0.9em}
	\label{tab:cam1}
	\setlength{\tabcolsep}{0.85mm}
	{
		\begin{tabular}{ccl}
			\hline
			\multicolumn{3}{c}{\textbf{\textit{Nikon-D3200} Specifications}}                                                         \\ \hline
			\multicolumn{1}{c|}{\textbf{Sensor Type}}        & \multicolumn{2}{c}{CMOS (Nikon DX format)}       \\
			\multicolumn{1}{c|}{\textbf{Sensor Size}}        & \multicolumn{2}{c}{23.2mm$\times$15.4mm} \\
			\multicolumn{1}{c|}{\textbf{Effective Pixels}}   & \multicolumn{2}{c}{24.7million}                  \\
			\multicolumn{1}{c|}{\textbf{Largest Image Size}} & \multicolumn{2}{c}{6016$\times$4000}             \\ \hline
		\end{tabular}
	}
	\vspace{-0.5em}
\end{table}

\begin{table}
	\renewcommand\arraystretch{1.3}
	\centering
	\scriptsize
	\vspace{-0.5em}
	\caption{Camera Captured Parameters of USTC-TD 2023}
	\vspace{-0.9em}
	\label{tab:cam2}
	\setlength{\tabcolsep}{0.85mm}
	{
		\begin{tabular}{ccl}
			\hline
			\multicolumn{3}{c}{\textbf{\textit{Nikon-Z-fc} Specifications}}                                                         \\ \hline
			\multicolumn{1}{c|}{\textbf{Sensor Type}}        & \multicolumn{2}{c}{CMOS (Nikon DX format)}       \\
			\multicolumn{1}{c|}{\textbf{Sensor Size}}        & \multicolumn{2}{c}{23.5mm$\times$15.7mm (APS-C)} \\
			\multicolumn{1}{c|}{\textbf{Effective Pixels}}   & \multicolumn{2}{c}{20.9million}                  \\
			\multicolumn{1}{c|}{\textbf{Largest Image Size}} & \multicolumn{2}{c}{5568$\times$3712}             \\ \hline
		\end{tabular}
	}
	\vspace{-1.5em}
\end{table}

\vspace{-0.3em}
\section{Data Collection}
In this section, we introduce the hardware, format and collection configuration of dataset collection.

\vspace{-1em}
\subsection{Camera and Format Configuration} 
\vspace{-0.2em}
The images and video sequences are captured by using \textit{Nikon-D3200} and \textit{Nikon-Z-fc} for USTC-TD 2022 and 2023 datasets, and the specific camera parameters are shown in Table~\ref{tab:cam1} and Table~\ref{tab:cam2}\,. For the format of images and video sequences in the dataset, they are transcoded from Raw camera format (DNG, MOV) and then converted to RGB, YUV444/420 color space/format by using the \textit{ffmpeg} tool and the conversion standard of color space (BT.601\cite{alshina2022ahg}).

\subsection{Collection Configuration}
\vspace{-0.1em}
To develop a comprehensive and diverse image/video dataset, we consider the various content factors of collected data, including the environmental factors (\textit{e.g.} scene type, texture, motion, view), imaging factors (\textit{e.g.} illumination, lens, shadow), which cover as close as possible to the real-world coding transmission scenes. For each factor, we categorize it into different types implicit in the construction of our dataset, the categories are shown in Table~\ref{tab:SC}\,. According to these conditions, we choose more than twenty different scene types (\textit{e.g.} dormitory, library, river bank, institutes, parks, classroom, street, vehicles), and adjust different camera parameters to capture. For each image, we take ten shots of the same scene at the same time to select the best one. For each video, we record five minutes for each scene with the same range of joint sense to select the short and long test sequences.

\begin{table}
	\renewcommand\arraystretch{1.15}
	\centering
	\scriptsize
	\caption{Collection Configuration of USTC-TD 2022 and 2023}
	\vspace{-0.9em}
	\label{tab:SC}
	\setlength{\tabcolsep}{2mm}
	{
		\begin{tabular}{ccc}
			\hline
			\multicolumn{3}{c}{\textbf{Collection Configuration}}                                                                                                                                                                                                   \\ \hline
			\multicolumn{1}{c|}{\textbf{Element}}  & \multicolumn{1}{c|}{\textbf{Category}}                                                           & \textbf{Example}                                                                                   \\ \hline
			\multicolumn{1}{c|}{\textbf{Texture}}      & \multicolumn{1}{c|}{\begin{tabular}[c]{@{}c@{}}Structural,\\ Natural,\\ Geometric\end{tabular}}  & \begin{tabular}[c]{@{}c@{}}Scenery,\\ People,\\ Gridding\end{tabular}                              \\ \hline
			\multicolumn{1}{c|}{\textbf{Motion}}       & \multicolumn{1}{c|}{\begin{tabular}[c]{@{}c@{}}Complex,\\ Medium,\\ Tiny\end{tabular}}           & \begin{tabular}[c]{@{}c@{}}Occlusion,\\ Walking,\\ Chatting\end{tabular}                           \\ \hline
			\multicolumn{1}{c|}{\textbf{View}}         & \multicolumn{1}{c|}{\begin{tabular}[c]{@{}c@{}}Upward Level,\\ Horizontal Level,\\ Overhead Level\end{tabular}}          & \begin{tabular}[c]{@{}c@{}}Building,\\ People,\\ Close Shot\end{tabular}                           \\ \hline
			\multicolumn{1}{c|}{\textbf{Illumination}} & \multicolumn{1}{c|}{\begin{tabular}[c]{@{}c@{}}Appropriate Exposure,\\ Underexposure,\\ Overexposure\end{tabular}} & \begin{tabular}[c]{@{}c@{}}Natural Light,\\ Dark Light,\\ High Light\end{tabular} \\ \hline
			\multicolumn{1}{c|}{\textbf{Lens}}         & \multicolumn{1}{c|}{\begin{tabular}[c]{@{}c@{}}Moving,\\ Fix\end{tabular}}          & \begin{tabular}[c]{@{}c@{}}Camera Motion,\\ Surveillance\end{tabular}                           \\ \hline
			\multicolumn{1}{c|}{\textbf{Shadow}}         & \multicolumn{1}{c|}{\begin{tabular}[c]{@{}c@{}}Hard, \\Soft, \\Cast\end{tabular}}          & \begin{tabular}[c]{@{}c@{}}Camera Flash Lamp,\\ Natural Illumination,\\ Building Occlusion\end{tabular}                           \\ \hline
		\end{tabular}
	\vspace{-1.5em}
	}
\end{table}

\vspace{-0.2em}
\section{Dataset Construction, Analysis, Discussion}
In this section, we introduce the construction of our proposed USTC-TD image and video datasets, and further analyze them based on the comparison with previous common test image/video datasets under different quantitative criteria.

\vspace{-0.8em}
\subsection{Construction of USTC-TD 2022 and 2023 Image Dataset}
\begin{figure*}
	\centering
	\vspace{-2em}
	\includegraphics[width=135mm]{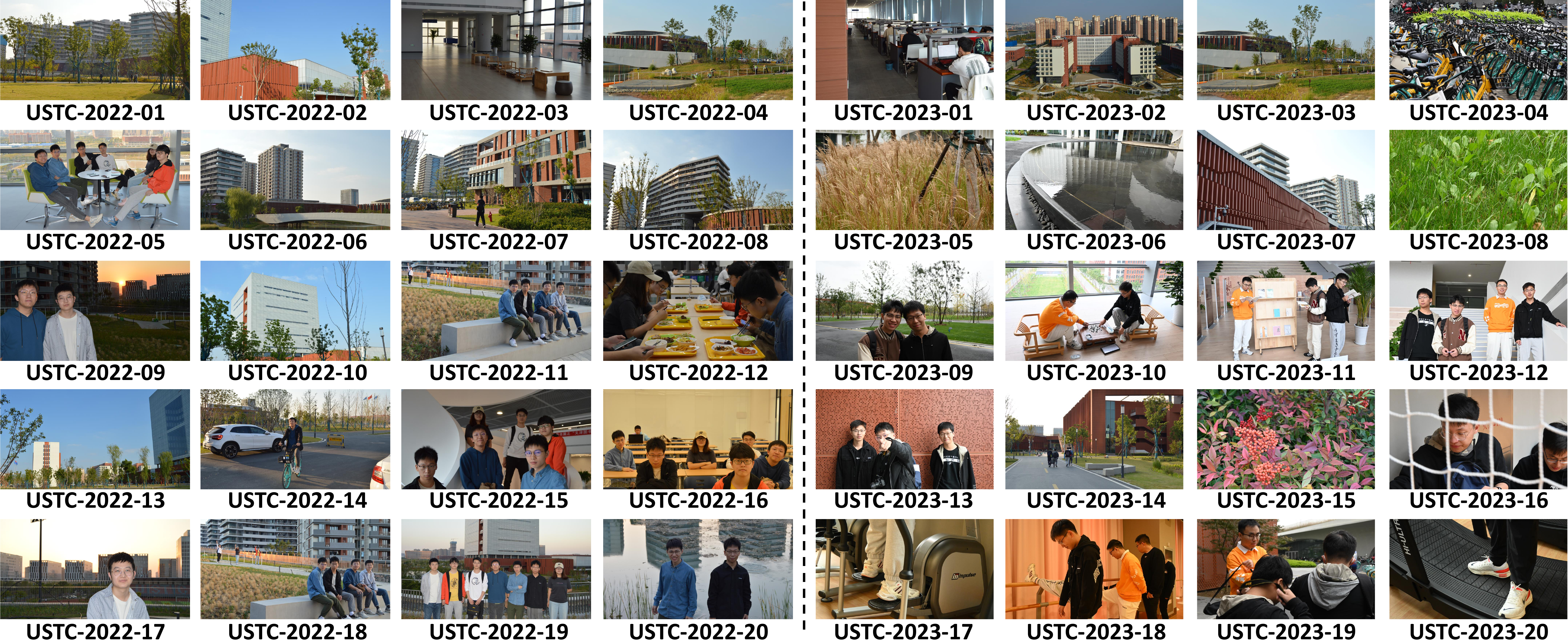}
	\vspace{-0.7em}
	\caption{Illustration of the image dataset in USTC-TD 2022 and 2023.}
	\vspace{-3em}
	\label{fig:USTC-TD-2022-1-crop}
	\label{fig:USTC-TD-2023-1-crop}
\end{figure*}

Based on the characteristics of previous image datasets\cite{Kodak, asuni2014testimages, clic}, our proposed dataset aims to cover various scenarios. Considering the various content factors, we combine different environmental factors and imaging factors in the collection process. For the diversity of environmental factors, we consider the scene type, texture, and view factors. For the diversity of imaging factors, we consider the resolution, illumination, and shadow factors. In Fig.\,~\ref{fig:USTC-TD-2022-1-crop}\,, we show all collected image data of USTD-TD 2022, and in Table~\ref{tab:ID2022}\,, we show the specific configuration of each image, and make it convenient for the researchers' scheme design for different application scenes. The collected image data and configuration of USTC-TD 2023 are also mentioned in Fig.\,~\ref{fig:USTC-TD-2023-1-crop}\, and Table~\ref{tab:ID2023}\,. Based on USTD-TD 2022, USTC-TD 2023 considers more extreme factors in real-world scenes.

\begin{table*}
	\renewcommand\arraystretch{1.0}
	\centering
	\fontsize{6pt}{8pt}\selectfont
	\vspace{-2em}
	\caption{The Configuration of USTC-TD 2022 Image Dataset}
	\vspace{-1.1em}
	\label{tab:ID2022}
	\setlength{\tabcolsep}{6mm}
	{
		\begin{tabular}{ccccccc}                                    \hline
			\multicolumn{1}{c|}{\textbf{Images}}       & \multicolumn{1}{c|}{\textbf{Scene Type}} & \multicolumn{1}{c|}{\textbf{Resolutions}}  & \multicolumn{1}{c|}{\textbf{Texture}} & \multicolumn{1}{c|}{\textbf{Illumination}}       & \multicolumn{1}{c|}{\textbf{View}} & \textbf{Shadow}        \\ \hline
			\multicolumn{1}{c|}{\textit{\textbf{USTC-2022-01}}} & \multicolumn{1}{c|}{Scenery}             & \multicolumn{1}{c|}{4096×2160}                   & \multicolumn{1}{c|}{Geometric}        & \multicolumn{1}{c|}{Appropriate Exposure}   &  \multicolumn{1}{c|}{Horizontal Level} & Soft \\
			\multicolumn{1}{c|}{\textit{\textbf{USTC-2022-02}}} & \multicolumn{1}{c|}{Scenery}             & \multicolumn{1}{c|}{4096×2160}                     & \multicolumn{1}{c|}{Structural}       & \multicolumn{1}{c|}{Appropriate Exposure}                          & \multicolumn{1}{c|}{Upward Level} & Soft     \\
			\multicolumn{1}{c|}{\textit{\textbf{USTC-2022-03}}} & \multicolumn{1}{c|}{Scenery}              & \multicolumn{1}{c|}{4096×2160}                            & \multicolumn{1}{c|}{Structural}       & \multicolumn{1}{c|}{Underexposure} & \multicolumn{1}{c|}{Horizontal Level}   & Hard                 \\
			\multicolumn{1}{c|}{\textit{\textbf{USTC-2022-04}}} & \multicolumn{1}{c|}{Scenery}             & \multicolumn{1}{c|}{4096×2160}                            & \multicolumn{1}{c|}{Structural}       & \multicolumn{1}{c|}{Appropriate Exposure}                          & \multicolumn{1}{c|}{Horizontal Level} & Soft                    \\
			\multicolumn{1}{c|}{\textit{\textbf{USTC-2022-05}}} & \multicolumn{1}{c|}{People}              & \multicolumn{1}{c|}{4096×2160}                    & \multicolumn{1}{c|}{Natural}          & \multicolumn{1}{c|}{Overexposure}           & \multicolumn{1}{c|}{Horizontal Level}   & Hard                         \\
			\multicolumn{1}{c|}{\textit{\textbf{USTC-2022-06}}} & \multicolumn{1}{c|}{Scenery}             & \multicolumn{1}{c|}{4096×2160}                        & \multicolumn{1}{c|}{Natural, Geometric}        & \multicolumn{1}{c|}{Appropriate Exposure}                          & \multicolumn{1}{c|}{Upward Level}    & Cast                    \\
			\multicolumn{1}{c|}{\textit{\textbf{USTC-2022-07}}} & \multicolumn{1}{c|}{Building}            & \multicolumn{1}{c|}{4096×2160}                         & \multicolumn{1}{c|}{Geometric}        & \multicolumn{1}{c|}{Appropriate Exposure}                          & \multicolumn{1}{c|}{Upward Level}  
			 & Cast                    \\
			\multicolumn{1}{c|}{\textit{\textbf{USTC-2022-08}}} & \multicolumn{1}{c|}{Scenery}             & \multicolumn{1}{c|}{4096×2160}                          & \multicolumn{1}{c|}{Geometric}        & \multicolumn{1}{c|}{Appropriate Exposure}                          & \multicolumn{1}{c|}{Upward Level}   & Cast                   \\
			\multicolumn{1}{c|}{\textit{\textbf{USTC-2022-09}}} & \multicolumn{1}{c|}{People, Building}    & \multicolumn{1}{c|}{4096×2160}                         & \multicolumn{1}{c|}{Natural}          & \multicolumn{1}{c|}{Underexposure}                          & \multicolumn{1}{c|}{Horizontal Level}    & Hard                   \\
			\multicolumn{1}{c|}{\textit{\textbf{USTC-2022-10}}} & \multicolumn{1}{c|}{Building}            & \multicolumn{1}{c|}{4096×2160}                   & \multicolumn{1}{c|}{Geometric}        & \multicolumn{1}{c|}{Appropriate Exposure}                          & \multicolumn{1}{c|}{Upward Level}  & Soft                           \\
			\multicolumn{1}{c|}{\textit{\textbf{USTC-2022-11}}} & \multicolumn{1}{c|}{People, Scenery}     & \multicolumn{1}{c|}{4096×2160}                          & \multicolumn{1}{c|}{Natural}          & \multicolumn{1}{c|}{Appropriate Exposure}                          & \multicolumn{1}{c|}{Horizontal Level}   & Soft                   \\
			\multicolumn{1}{c|}{\textit{\textbf{USTC-2022-12}}} & \multicolumn{1}{c|}{People}              & \multicolumn{1}{c|}{4096×2160}                            & \multicolumn{1}{c|}{Natural}          & \multicolumn{1}{c|}{Underexposure}                          & \multicolumn{1}{c|}{Horizontal Level}   & Hard                 \\
			\multicolumn{1}{c|}{\textit{\textbf{USTC-2022-13}}} & \multicolumn{1}{c|}{Scenery}             & \multicolumn{1}{c|}{4096×2160}                           & \multicolumn{1}{c|}{Nature, Structural}       & \multicolumn{1}{c|}{Appropriate Exposure}                          & \multicolumn{1}{c|}{Upward Level} & Cast                    \\
			\multicolumn{1}{c|}{\textit{\textbf{USTC-2022-14}}} & \multicolumn{1}{c|}{People, Vehicle}     & \multicolumn{1}{c|}{4096×2160}                          & \multicolumn{1}{c|}{Natural}          & \multicolumn{1}{c|}{Underexposure}                          & \multicolumn{1}{c|}{Horizontal Level}  & Cast                    \\
			\multicolumn{1}{c|}{\textit{\textbf{USTC-2022-15}}} & \multicolumn{1}{c|}{People}              & \multicolumn{1}{c|}{4096×2160}                           & \multicolumn{1}{c|}{Natural}          & \multicolumn{1}{c|}{Underexposure}                          & \multicolumn{1}{c|}{Upward Level}    & Hard                 \\
			\multicolumn{1}{c|}{\textit{\textbf{USTC-2022-16}}} & \multicolumn{1}{c|}{People}              & \multicolumn{1}{c|}{4096×2160}                         & \multicolumn{1}{c|}{Natural}          & \multicolumn{1}{c|}{Appropriate Exposure}                          & \multicolumn{1}{c|}{Horizontal Level}     & Soft                  \\
			\multicolumn{1}{c|}{\textit{\textbf{USTC-2022-17}}} & \multicolumn{1}{c|}{People, Building}    & \multicolumn{1}{c|}{4096×2160}                          & \multicolumn{1}{c|}{Natural}          & \multicolumn{1}{c|}{Overexposure}                          & \multicolumn{1}{c|}{Horizontal Level}   & Hard                   \\
			\multicolumn{1}{c|}{\textit{\textbf{USTC-2022-18}}} & \multicolumn{1}{c|}{People, Building}    & \multicolumn{1}{c|}{4096×2160}                           & \multicolumn{1}{c|}{Natural, Geometric}          & \multicolumn{1}{c|}{Appropriate Exposure}                          & \multicolumn{1}{c|}{Horizontal Level}   & Soft                  \\
			\multicolumn{1}{c|}{\textit{\textbf{USTC-2022-19}}} & \multicolumn{1}{c|}{People, Building}    & \multicolumn{1}{c|}{4096×2160}                     & \multicolumn{1}{c|}{Natural, Geometric}          & \multicolumn{1}{c|}{Overexposure}                          & \multicolumn{1}{c|}{Horizontal Level}   & Soft                        \\
			\multicolumn{1}{c|}{\textit{\textbf{USTC-2022-20}}} & \multicolumn{1}{c|}{People, River}              & \multicolumn{1}{c|}{4096×2160}                         & \multicolumn{1}{c|}{Natural}          & \multicolumn{1}{c|}{Appropriate Exposure}                          & \multicolumn{1}{c|}{Overhead Level}   & Cast   \\ \hline
		\end{tabular}
	}
	\vspace{1em}
	\renewcommand\arraystretch{1.0}
	\centering
	\fontsize{6pt}{8pt}\selectfont
	\caption{The Configuration of USTC-TD 2023 Image Dataset}
	\vspace{-1.1em}
	\label{tab:ID2023}
	\setlength{\tabcolsep}{5.5mm}
	{
		\begin{tabular}{ccccccc}
			\hline
			\multicolumn{1}{c|}{\textbf{Images}}       & \multicolumn{1}{c|}{\textbf{Scene Type}}    & \multicolumn{1}{c|}{\textbf{Resolutions}}  & \multicolumn{1}{c|}{\textbf{Texture}} & \multicolumn{1}{c|}{\textbf{Illumination}}       & \multicolumn{1}{c|}{\textbf{View}} & \textbf{Shadow}        \\ \hline
			\multicolumn{1}{c|}{\textit{\textbf{USTC-2023-01}}} & \multicolumn{1}{c|}{People, Room}           & \multicolumn{1}{c|}{3840×2160}                     & \multicolumn{1}{c|}{Natural, Structural}          & \multicolumn{1}{c|}{Appropriate Exposure}   & \multicolumn{1}{c|}{Horizontal Level} & Hard \\
			\multicolumn{1}{c|}{\textit{\textbf{USTC-2023-02}}} & \multicolumn{1}{c|}{Scenery}                & \multicolumn{1}{c|}{3840×2160}                           & \multicolumn{1}{c|}{Geometric, Structural}        & \multicolumn{1}{c|}{Underexposure} & \multicolumn{1}{c|}{Overhead Level} & Cast    \\
			\multicolumn{1}{c|}{\textit{\textbf{USTC-2023-03}}} & \multicolumn{1}{c|}{Scenery}                & \multicolumn{1}{c|}{3840×2160}                           & \multicolumn{1}{c|}{Natural, Structural}       & \multicolumn{1}{c|}{Underexposure}                          & \multicolumn{1}{c|}{Horizontal Level} & Soft                    \\
			\multicolumn{1}{c|}{\textit{\textbf{USTC-2023-04}}} & \multicolumn{1}{c|}{Bicycle, Dense Objects} & \multicolumn{1}{c|}{3840×2160}                           & \multicolumn{1}{c|}{Geometric}        & \multicolumn{1}{c|}{Appropriate Exposure}                          & \multicolumn{1}{c|}{Horizontal Level}  & Soft                   \\
			\multicolumn{1}{c|}{\textit{\textbf{USTC-2023-05}}} & \multicolumn{1}{c|}{Plant, Dense Textures}  & \multicolumn{1}{c|}{3840×2160}                     & \multicolumn{1}{c|}{Structural}       & \multicolumn{1}{c|}{Overexposure}           & \multicolumn{1}{c|}{Horizontal Level}   & Soft                        \\
			\multicolumn{1}{c|}{\textit{\textbf{USTC-2023-06}}} & \multicolumn{1}{c|}{Water Wave}             & \multicolumn{1}{c|}{3840×2160}                           & \multicolumn{1}{c|}{Geometric}        & \multicolumn{1}{c|}{Appropriate Exposure}                          & \multicolumn{1}{c|}{Overhead Level}    & Cast                 \\
			\multicolumn{1}{c|}{\textit{\textbf{USTC-2023-07}}} & \multicolumn{1}{c|}{Building}               & \multicolumn{1}{c|}{3840×2160}                           & \multicolumn{1}{c|}{Geometric}        & \multicolumn{1}{c|}{Overexposure}                          & \multicolumn{1}{c|}{Upward Level} & Soft     \\
			\multicolumn{1}{c|}{\textit{\textbf{USTC-2023-08}}} & \multicolumn{1}{c|}{Plant, Dense Textures}  & \multicolumn{1}{c|}{3840×2160}                            & \multicolumn{1}{c|}{Structural}       & \multicolumn{1}{c|}{Appropriate Exposure}                          & \multicolumn{1}{c|}{Overhead Level}   & Soft                 \\
			\multicolumn{1}{c|}{\textit{\textbf{USTC-2023-09}}} & \multicolumn{1}{c|}{Scenery, People}        & \multicolumn{1}{c|}{3840×2160}                            & \multicolumn{1}{c|}{Natural}          & \multicolumn{1}{c|}{Underexposure}                          & \multicolumn{1}{c|}{Horizontal Level}  & Hard                  \\
			\multicolumn{1}{c|}{\textit{\textbf{USTC-2023-10}}} & \multicolumn{1}{c|}{People}         & \multicolumn{1}{c|}{3840×2160}                            & \multicolumn{1}{c|}{Natural}          & \multicolumn{1}{c|}{Overexposure}                          & \multicolumn{1}{c|}{Overhead Level}    & Soft                \\
			\multicolumn{1}{c|}{\textit{\textbf{USTC-2023-11}}} & \multicolumn{1}{c|}{People}         & \multicolumn{1}{c|}{3840×2160}                            & \multicolumn{1}{c|}{Natural}          & \multicolumn{1}{c|}{Overexposure}                          & \multicolumn{1}{c|}{Horizontal Level}   & Soft                 \\
			\multicolumn{1}{c|}{\textit{\textbf{USTC-2023-12}}} & \multicolumn{1}{c|}{People}                 & \multicolumn{1}{c|}{3840×2160}                           & \multicolumn{1}{c|}{Natural}          & \multicolumn{1}{c|}{Underexposure}                          & \multicolumn{1}{c|}{Upward Level}    & Soft                 \\
			\multicolumn{1}{c|}{\textit{\textbf{USTC-2023-13}}} & \multicolumn{1}{c|}{People}                 & \multicolumn{1}{c|}{3840×2160}                            & \multicolumn{1}{c|}{Natural}          & \multicolumn{1}{c|}{Appropriate Exposure}                          & \multicolumn{1}{c|}{Horizontal Level}  & Soft                  \\
			\multicolumn{1}{c|}{\textit{\textbf{USTC-2023-14}}} & \multicolumn{1}{c|}{Building}               & \multicolumn{1}{c|}{3840×2160}                            & \multicolumn{1}{c|}{Geometric}        & \multicolumn{1}{c|}{Appropriate Exposure}                          & \multicolumn{1}{c|}{Horizontal Level}   & Cast                 \\
			\multicolumn{1}{c|}{\textit{\textbf{USTC-2023-15}}} & \multicolumn{1}{c|}{Plant}                  & \multicolumn{1}{c|}{3840×2160}                           & \multicolumn{1}{c|}{Structural}       & \multicolumn{1}{c|}{Appropriate Exposure}                          & \multicolumn{1}{c|}{Horizontal Level}  & Soft                   \\
			\multicolumn{1}{c|}{\textit{\textbf{USTC-2023-16}}} & \multicolumn{1}{c|}{People, Occlusion}      & \multicolumn{1}{c|}{3840×2160}                           & \multicolumn{1}{c|}{Natural}          & \multicolumn{1}{c|}{Appropriate Exposure}                          & \multicolumn{1}{c|}{Horizontal Level}    & Soft                 \\
			\multicolumn{1}{c|}{\textit{\textbf{USTC-2023-17}}} & \multicolumn{1}{c|}{Close Shot}             & \multicolumn{1}{c|}{3840×2160}                           & \multicolumn{1}{c|}{Natural}          & \multicolumn{1}{c|}{Appropriate Exposure}                          & \multicolumn{1}{c|}{Horizontal Level}   & Soft                  \\
			\multicolumn{1}{c|}{\textit{\textbf{USTC-2023-18}}} & \multicolumn{1}{c|}{People}                 & \multicolumn{1}{c|}{3840×2160}                            & \multicolumn{1}{c|}{Natural}          & \multicolumn{1}{c|}{Appropriate Exposure}                          & \multicolumn{1}{c|}{Horizontal Level} & Soft                   \\
			\multicolumn{1}{c|}{\textit{\textbf{USTC-2023-19}}} & \multicolumn{1}{c|}{Close Shot}             & \multicolumn{1}{c|}{3840×2160}                   & \multicolumn{1}{c|}{Natural}          & \multicolumn{1}{c|}{Underexposure}                          & \multicolumn{1}{c|}{Horizontal Level}   & Hard                 \\
			\multicolumn{1}{c|}{\textit{\textbf{USTC-2023-20}}} & \multicolumn{1}{c|}{Close Shot}             & \multicolumn{1}{c|}{3840×2160}                            & \multicolumn{1}{c|}{Natural}          & \multicolumn{1}{c|}{Appropriate Exposure}                          & \multicolumn{1}{c|}{Overhead Level}    & Cast                \\ \hline
		\end{tabular}
	}
	\vspace{1.1em}
	\renewcommand\arraystretch{1.25}
	\centering
	\fontsize{6pt}{8pt}\selectfont
	\caption{The Configuration of USTC-TD 2023 Video Dataset}
	\vspace{-1.1em}
	\label{tab:VD2023}
	\setlength{\tabcolsep}{3.4mm}
	{
		\begin{tabular}{ccccccccc}
			\hline
			\multicolumn{1}{c|}{\textbf{Video Sequences}}      & \multicolumn{1}{c|}{\textbf{Color Space}} & \multicolumn{1}{c|}{\textbf{Motion}}                & \multicolumn{1}{c|}{\textbf{Scene Types}} & \multicolumn{1}{c|}{\textbf{Resolutions}} & \multicolumn{1}{c|}{\textbf{Quality}} & \multicolumn{1}{c|}{\textbf{Texture}} & \multicolumn{1}{c|}{\textbf{View}}         & \textbf{Lens} \\ \hline
			\multicolumn{1}{c|}{\textbf{\textit{USTC-Badminton}}}       & \multicolumn{1}{l|}{YUV420, 444, RGB}     & \multicolumn{1}{c|}{{\color[HTML]{6434FC} \textbf{Medium}}}  & \multicolumn{1}{c|}{People, Sport}        & \multicolumn{1}{c|}{1920×1080}            & \multicolumn{1}{c|}{High}       & \multicolumn{1}{c|}{Natural}          & \multicolumn{1}{c|}{Horizontal Level} & Moving          \\
			\multicolumn{1}{c|}{\textbf{\textit{USTC-BasketballDrill}}} & \multicolumn{1}{l|}{YUV420, 444, RGB}     & \multicolumn{1}{c|}{{\color[HTML]{6434FC} \textbf{Medium}}}  & \multicolumn{1}{c|}{People, Sport}        & \multicolumn{1}{c|}{1920×1080}            & \multicolumn{1}{c|}{High}         & \multicolumn{1}{c|}{Natural, Geometric}          & \multicolumn{1}{c|}{Horizontal Level}                    & Moving          \\
			\multicolumn{1}{c|}{\textbf{\textit{USTC-BasketballPass}}}  & \multicolumn{1}{l|}{YUV420, 444, RGB}     & \multicolumn{1}{c|}{{\color[HTML]{6434FC} \textbf{Medium}}}  & \multicolumn{1}{c|}{People, Sport}        & \multicolumn{1}{c|}{1920×1080}            & \multicolumn{1}{c|}{High}                & \multicolumn{1}{c|}{Natural, Geometric}          & \multicolumn{1}{c|}{Horizontal Level}                    & Moving          \\
			\multicolumn{1}{c|}{\textbf{\textit{USTC-BicycleDriving}}}  & \multicolumn{1}{l|}{YUV420, 444, RGB}     & \multicolumn{1}{l|}{{\color[HTML]{FE0000} \textbf{Complex}}} & \multicolumn{1}{c|}{People, Daily Life}   & \multicolumn{1}{c|}{1920×1080}            & \multicolumn{1}{c|}{High}                & \multicolumn{1}{c|}{Natural, Structural}          & \multicolumn{1}{c|}{Horizontal Level}                    & Moving          \\
			\multicolumn{1}{c|}{\textbf{\textit{USTC-Dancing}}}         & \multicolumn{1}{l|}{YUV420, 444, RGB}     & \multicolumn{1}{l|}{{\color[HTML]{FE0000} \textbf{Complex}}} & \multicolumn{1}{c|}{People, Sport}        & \multicolumn{1}{c|}{1920×1080}            & \multicolumn{1}{c|}{High}                & \multicolumn{1}{c|}{Natural}          & \multicolumn{1}{c|}{Horizontal Level}                    & Fix           \\
			\multicolumn{1}{c|}{\textbf{\textit{USTC-ParkWalking}}}     & \multicolumn{1}{l|}{YUV420, 444, RGB}     & \multicolumn{1}{l|}{{\color[HTML]{FE0000} \textbf{Complex}}} & \multicolumn{1}{c|}{People, Daily Life}   & \multicolumn{1}{c|}{1920×1080}            & \multicolumn{1}{c|}{High}                & \multicolumn{1}{c|}{Natural, Structural}          & \multicolumn{1}{c|}{Horizontal Level}                    & Moving          \\
			\multicolumn{1}{c|}{\textbf{\textit{USTC-Running}}  }       & \multicolumn{1}{l|}{YUV420, 444, RGB}     & \multicolumn{1}{l|}{{\color[HTML]{FE0000} \textbf{Complex}}} & \multicolumn{1}{c|}{People, Sport}        & \multicolumn{1}{c|}{1920×1080}            & \multicolumn{1}{c|}{High}                & \multicolumn{1}{c|}{Natural, Structural}          & \multicolumn{1}{c|}{Horizontal Level}                    & Moving          \\
			\multicolumn{1}{c|}{\textbf{\textit{USTC-ShakingHands} }}   & \multicolumn{1}{l|}{YUV420, 444, RGB}     & \multicolumn{1}{l|}{{\color[HTML]{FE0000} \textbf{Complex}}} & \multicolumn{1}{l|}{People, Daliy life}   & \multicolumn{1}{c|}{1920×1080}            & \multicolumn{1}{c|}{High}                & \multicolumn{1}{c|}{Natural, Geometric}          & \multicolumn{1}{c|}{Horizontal Level}                    & Moving          \\
			\multicolumn{1}{c|}{\textbf{\textit{USTC-Snooker}} }        & \multicolumn{1}{l|}{YUV420, 444, RGB}     & \multicolumn{1}{c|}{{\color[HTML]{3531FF} \textbf{Tiny}}}    & \multicolumn{1}{c|}{Sport}                & \multicolumn{1}{c|}{1920×1080}            & \multicolumn{1}{c|}{High}                & \multicolumn{1}{c|}{Natural}          & \multicolumn{1}{c|}{Horizontal Level}                    & Moving          \\
			\multicolumn{1}{c|}{\textbf{\textit{USTC-FourPeople} }}     & \multicolumn{1}{l|}{YUV420, 444, RGB}                          & \multicolumn{1}{c|}{{\color[HTML]{3531FF} \textbf{Tiny}}}    & \multicolumn{1}{c|}{People}               & \multicolumn{1}{c|}{1920×1080}            & \multicolumn{1}{c|}{High}                & \multicolumn{1}{c|}{Natural}          & \multicolumn{1}{c|}{Horizontal Level}                    & Fix           \\ \hline
		\end{tabular}
		\vspace{-1.3em}
	}
\end{table*}

\begin{figure*}
	\centering
	\vspace{-1.2em}
	\includegraphics[width=120mm]{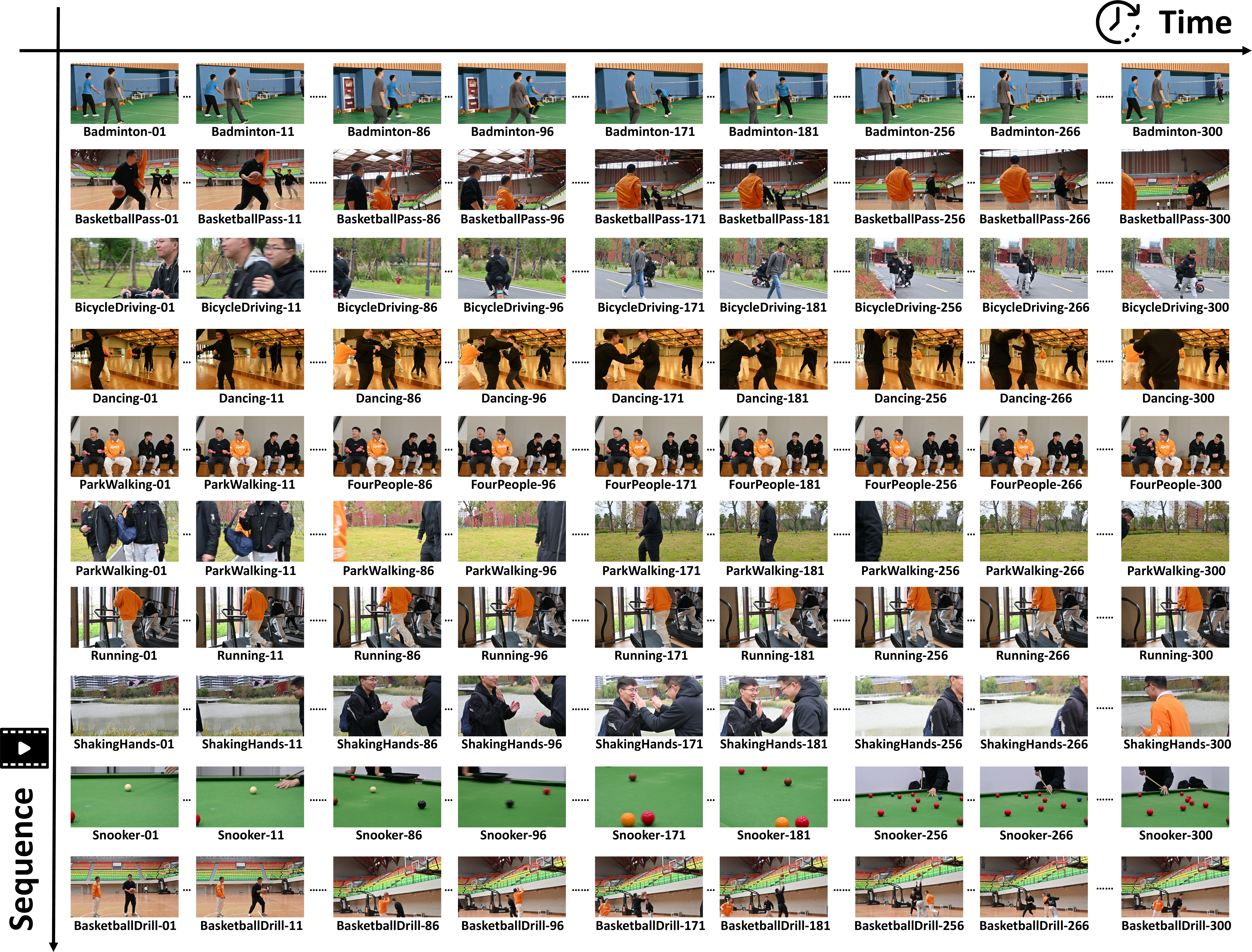}
	\vspace{-0.7em}
	\caption{Illustration of each video sequence in USTC-TD 2023 video dataset. The 0$\sim$96/0$\sim$300 frames correspond to the short and long setting.}
	\label{fig:USTC-TD-2023V-crop}
	\vspace{-1em}
\end{figure*}

Compared to the previous image datasets\cite{Kodak, asuni2014testimages, clic}, more specific content factors are considered in our dataset. For example, in \textit{USTC-2022-09} and \textit{USTC-2022-05}, we capture the low-light image with underexposure and high-light image with overexposure, which is a challenge for the generalization of many researchers' image compression schemes \cite{shen2023dec, cai2024make}. 
In \textit{USTC-2023-16}, we capture the scenes with the object occlusion and the spatial-wise correlation becomes low, which is a challenge for traditional intra-prediction schemes \cite{pfaff2021intra} in the traditional codec\cite{sullivan2012overview,bross2021overview,ma2022evolution,han2021technical,zhao2021study}. We hope these specific testing sets can help the researchers discover the problem related to spatial characteristics in their image compression scheme.

\subsection{Construction of USTC-TD 2023 Video Dataset}
\vspace{-0.3em}
Based on the characteristics of previous video datasets\cite{mercat2020uvg, wang2016mcl, bossen2013common, boyce2018jvet}, our proposed dataset aims to cover more typical characteristics of video content. Compared to the image data, temporal-domain properties are unique to video, especially in the diverse motion types with more environmental and imaging factors in natural videos. There are usually multiple moving objects of arbitrary shapes and various motion types in video frames, leading to complex motion fields, which challenge the video coding schemes\cite{li2017efficient, li2024object, li2024uniformly, li2024loop}. Therefore, we simulate the video data with various temporal correlation types, including different kinds of motion types and lens motion. 

In Fig.~\ref{fig:USTC-TD-2023V-crop}\,, we show the partial frames of all collected video sequences in the USTD-TD 2023 dataset, and the specific configuration of each video in Table~\ref{tab:VD2023}\,, which make it convenient for the scheme design of research in real-world scenes under different temporal correlation types. Note that the different motion types are obviously classified in the Table~\ref{tab:VD2023}\, with different colors. For the construction of USTC-TD video dataset, two kinds of settings are constructed: short setting and long setting. For the short setting (3 seconds), a 96-frame subset is selected from each full-length captured test video sequence (frame rate: 30fps, 300 seconds, 9000 frames in total) to reduce the high complexity associated with long sequences during the practical testing, which can promote the fast evaluation in the research process. For the long setting, we extend from the last frame of the short setting and directly enlarge the short sequence to 300 frames (10 seconds), which aligns with the 10-second setting of test sequence length of HEVC and VVC common test condition (CTC).

Different from the previous video dataset, we add more specific temporal correlation types in our proposed video dataset. For example, in \textit{USTC-BicycleDriving}, we capture the video with a fast scene change (lens motion), high-speed moving objects, and object occlusion, which is a challenge for many inter-alignment schemes of submitted solutions in \textit{VCIP Challenge} 2023\,\footnote{Available online at \,\url{https://vcip2023.iforum.biz/page/goto/}.} and many optical flow-based video compression schemes \cite{lu2019dvc, lu2020end,ho2022canf, li2021deep, sheng2022temporal,li2022hybrid, tang2024offline,sheng2024spatial,sheng2024vnvc,li2023neural,li2024neural} (the detailed performance analysis is mentioned in Section V.B). For the performance of different schemes of this sequence, the learned video compression schemes are far inferior to the traditional codecs \cite{sullivan2012overview, bross2021overview, ma2022evolution, han2021technical, zhao2021study}. In \textit{USTC-Snooker}, we capture the scenes with the tiny motion and fast lens motion, which is also a challenge for optical flow-based schemes. At present, most of the learned video compression schemes use the optical flow-based alignment\cite{sun2018pwc, ranjan2017optical}, and the optical flow-based motion estimation is difficult to capture the tiny motion and further influences their performance. Therefore, we put forward our proposed video dataset with the above specific designs, and hope the efficient testing datasets can help the researchers discover the problem related to temporal characteristics on their video compression scheme.

\begin{figure}
	\centering
	\includegraphics[width=80mm]{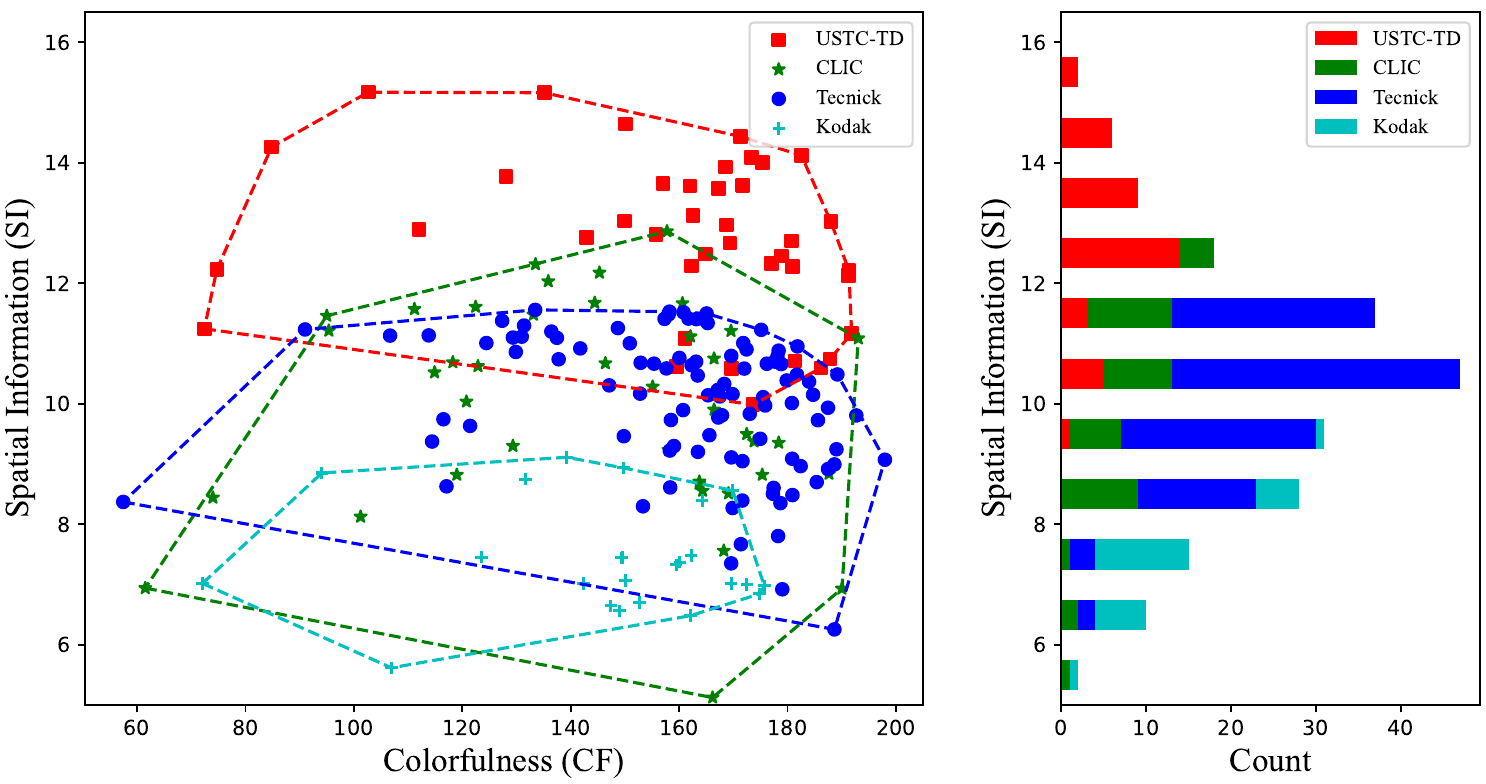}
	\vspace{-0.5em}
	\caption{The visualization of the evaluation of spatial information (SI) and colorfulness (CF) features on different image test datasets. Scatter diagram represents the SI versus CF, and corresponding convex hulls indicate the coverage of different datasets. The histogram represents the number of images under different SI scores.}
	\label{fig:Image_CF_SI}
	
	\vspace{0.9em}
	\centering
	\includegraphics[width=80mm]{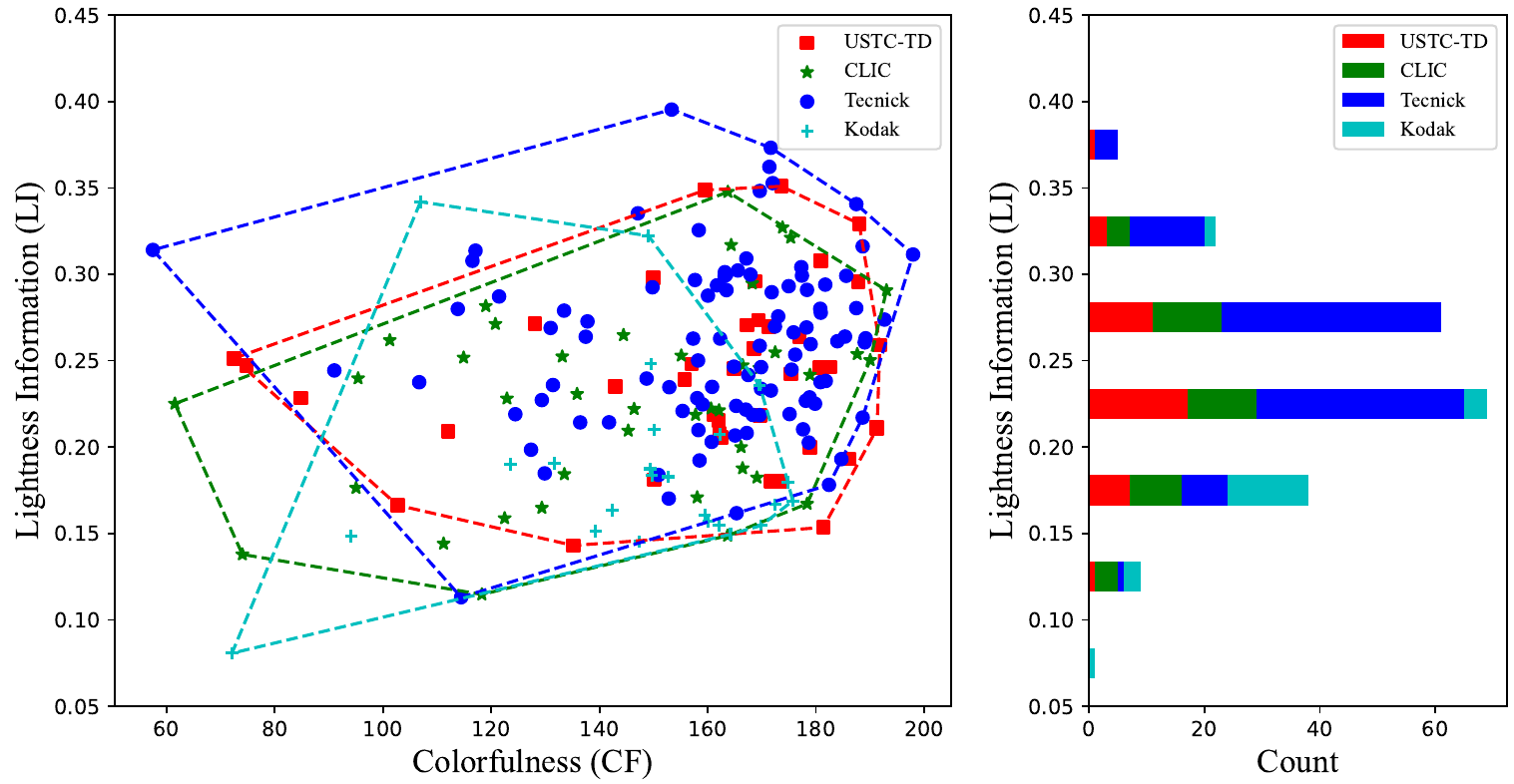}
	\vspace{-0.9em}
	\caption{The visualization of the evaluation of lightness information (LI) and CF features on different image test datasets. Scatter diagram represents LI versus CF, and corresponding convex hulls indicate the coverage of different datasets. The histogram represents the number of images under different LI scores.}
	\label{fig:light_SI_LI}
	
	\vspace{0.9em}
	\centering
	\includegraphics[width=80mm]{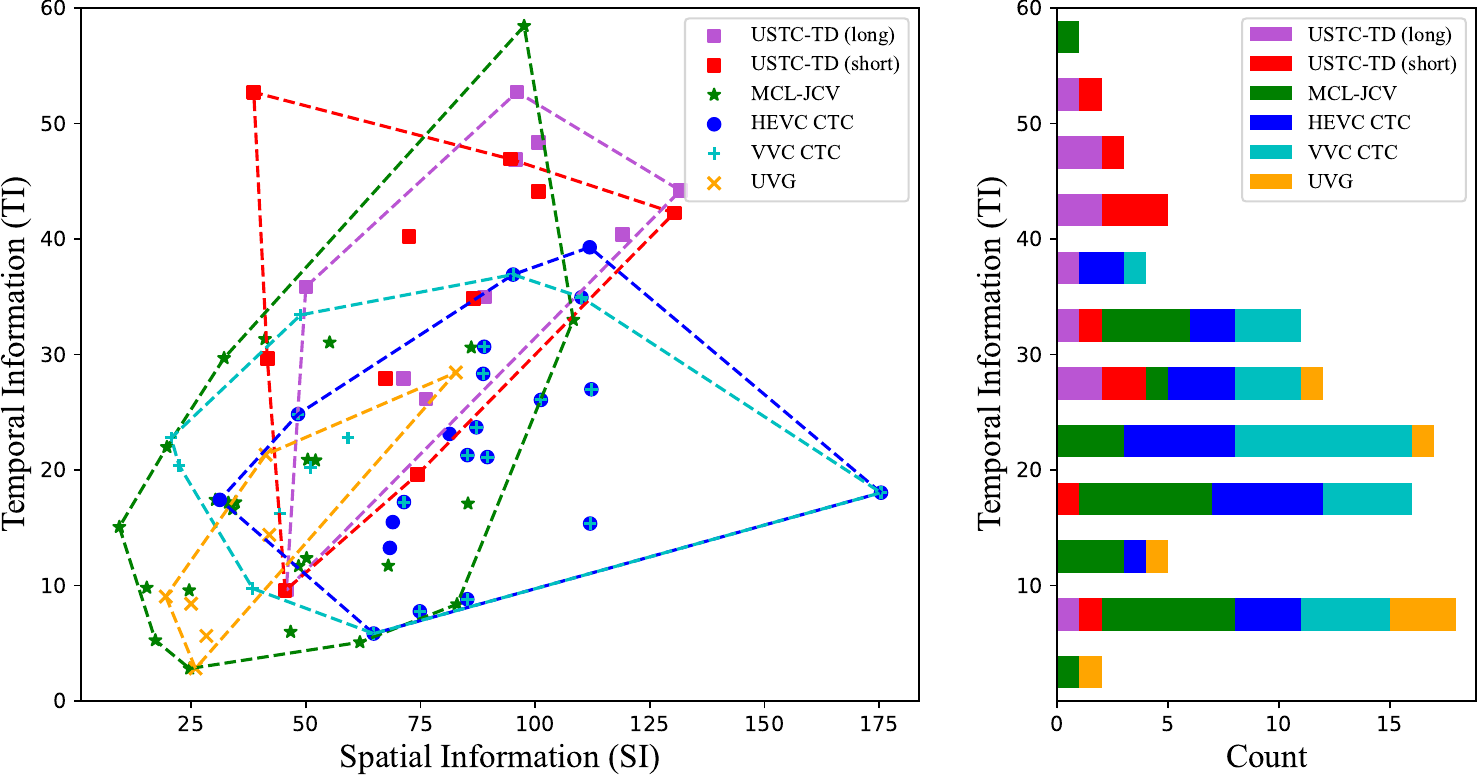}
	\vspace{-0.8em}
	\caption{The visualization of the evaluation of  temporal information (TI) and SI features on different video test datasets. Scatter diagram represents the TI versus SI, and corresponding convex hulls indicate the coverage of different datasets. The histogram represents the number of videos under different TI scores.}
	\label{fig:video_SI_TI}
	\vspace{-1.3em}
\end{figure}

\vspace{-0.8em}
\subsection{Analysis of USTC-TD Image and Video datasets}
In the above subsections, the construction of each dataset of USTC-TD is introduced, here we point out the specific characteristics of the proposed dataset and highlight the driving factors behind its benefit, and discuss its practical application.

\subsubsection{Characteristics of USTC-TD Image and Video datasets}
To comprehensively verify the outstanding coverage of our proposed dataset for various content factors and qualitatively analyze the superiority of USTC-TD, we evaluate the USTC-TD on different image/video features and compare it with the previous image/video common test datasets (image datasets: \textit{Kodak}\cite{Kodak}, \textit{CLIC}\cite{clic}, \textit{Tecnick}\cite{asuni2014testimages}, video datasets: \textit{HEVC CTC}\cite{bossen2013common}, \textit{VVC CTC}\cite{boyce2018jvet}, \textit{MCL-JCV}\cite{wang2016mcl}, \textit{UVG}\cite{mercat2020uvg}). For analysis of image/video features, we select the spatial information (SI)\cite{winkler2012analysis}, colorfulness (CF)\cite{winkler2012analysis},  lightness information (LI)\cite{wang2013naturalness}, and temporal information (TI)\cite{mercat2020uvg} to characterize each dataset along the dimensions of space, color, lightness, and temporal correlation, which are commonly used to evaluate the quality of dataset \cite{ma2021bvi, mackin2018study, mercat2020uvg}. The definitions of the evaluation of these features can be found in \cite{winkler2012analysis, mercat2020uvg, itu1999subjective, wang2013naturalness, peli1990contrast}, and the detailed ways are as follows:

\begin{itemize}
\item \textit{\textbf{Spatial information (SI)}}: SI is used as a representation of edge energy\cite{standard2003digital}. Followed by \cite{winkler2012analysis}, the SI is defined as the root mean square of edge magnitude over the luma component of an image or a video frame:
\begin{equation}
    Score_{SI}= \sqrt{\frac{L}{1080}}\sqrt{\sum{\frac{S_r^2}{P}}},
\end{equation}
where $S_{r} = \sqrt{S_{v}^{2}+S_{h}^{2}}$ indicates the edge magnitude at each pixel. $S_{v}$ and $S_{h}$ indicate the images/video frames filtered with vertical and horizontal \textit{Sobel} kernels, respectively. $P$ indicates the total number of pixels in the filtered image, and $L$ indicates the vertical resolution. The normalization factor $\sqrt{\frac{L}{1080}}$ is used to reduce the scale and resolution dependence of SI. For video datasets, followed by \cite{mercat2020uvg}, SI is taking the maximum of the results of all video frames.

\item \textit{\textbf{Colorfulness (CF)}}: CF is used as a representation of the variety and intensity of colors in the image. Followed by \cite{hasler2003measuring, mercat2020uvg}, CF is defined as 
\begin{equation}
    Score_{CF} = \sqrt{\sigma_{rg}^2+\sigma_{by}^2}+0.3\sqrt{\mu_{rg}^2+\mu_{by}^2}
\end{equation}
where opponent color spaces ($rg$, $by$) are defined in RGB color space. To be special, $rg = R-G$ and $by = 0.5\,(R+G)-B$.

\item \textit{\textbf{Lightness information (LI)}}: LI is used as a representation of lightness variation. To measure the lightness information, we adopt the root mean square (RMS) contrast\cite{peli1990contrast}, the LI is defined as the standard deviation of the pixel intensities:

\begin{equation}
Score_{LI} = \sqrt{\frac{1}{MN}\sum_{i=0}^{N-1}\sum_{j=0}^{M-1}(I_{i,j}-\hat{I})^2},
\end{equation}
where the intensities ($I_{i,j}$) are the $i$-th and $j$-th elements of the two-dimensional image of size $M$ by $N$. $\hat{I}$ is the average intensity of all pixel values in the image. The pixel intensities of the image ($I$) are normalized in the range $[0,1]$.

\item \textit{\textbf{Temporal information (TI)}}: TI is used as a representation of temporal variation. Followed by \cite{itu1999subjective, mercat2020uvg}, TI is defined as the maximum amount of temporal variation between successive frames $F_{n-1}$ and $F_{n}$:
\begin{equation}
\hspace{-0.3em} Score_{TI}=\max_{1\leq n\leq N-1}\left\{\underset{\substack{0\leq i\leq W-1 \\  0\leq j\leq H-1}}{\mathrm{std}}\left[F_n(i,j) - F_{n-1}(i,j)\right]\right\}
\end{equation}
where $W$, $H$, $N$ denote the frame width, height and the number of total frames, respectively.
\end{itemize}

The quantitative evaluation scores of different datasets are shown in Fig.\,\ref{fig:Image_CF_SI}\,,\,\,\ref{fig:light_SI_LI}\,,\,\,\ref{fig:video_SI_TI}\,. From the comparison with other datasets, we find that USTC-TD can collaborate with other datasets to handle a wide coverage of different image/video features, which verifies the diversity of the proposed dataset. 

\begin{table}
	\renewcommand\arraystretch{0.95}
	\centering
	\scriptsize
	\caption{Quantitative Results of the USTC-TD 2022 and 2023 Image Datasets. Note the Higher Scores are Represented in \textcolor{red}{Red},\\ and the Lower Scores are Represented in \textcolor{blue}{Blue}}
	\vspace{-0.6em}
	\label{tab:image_metrics}
	\setlength{\tabcolsep}{2.0mm}
	{
		\begin{tabular}{cclcclcc}
			\toprule
			\multicolumn{4}{c|}{\textbf{\textit{USTC-TD 2022}}}                                                                                    & \multicolumn{4}{c}{\textbf{\textit{USTC-TD 2023}}}                                     \\ \midrule
			\multicolumn{1}{c|}{\textbf{\textit{Image}}} & \textbf{\textit{SI}}    & \textbf{\textit{\,\,\,\,CF}}     & \textbf{\textit{LI}}   & \multicolumn{1}{|c|}{\textbf{\textit{Image}}} & \textbf{\textit{\,\,\,\,SI}}    & \textbf{\textit{CF}}     & \textbf{\textit{LI}}   \\
			\midrule
			\multicolumn{1}{c|}{\textbf{\textit{01}}} & \textcolor{red}{13.57} & 167.24 & \textcolor{red}{0.27} & \multicolumn{1}{|c|}{\textbf{\textit{01}}} & 12.81 & 155.76 & 0.24 \\
			\multicolumn{1}{c|}{\textbf{\textit{02}}} & \textcolor{blue}{10.71} & 181.37 & \textcolor{blue}{0.15} & \multicolumn{1}{|c|}{\textbf{\textit{02}}} & 13.13 & 162.53 & 0.21 \\
			\multicolumn{1}{c|}{\textbf{\textit{03}}} & 11.08 & 161.07 & 0.22 & \multicolumn{1}{|c|}{\textbf{\textit{03}}} & \textcolor{blue}{12.22} & \textcolor{red}{191.29} & 0.21 \\
			\multicolumn{1}{c|}{\textbf{\textit{04}}} & 12.13 & \textcolor{red}{191.26} & 0.21 & \multicolumn{1}{|c|}{\textbf{\textit{04}}} & \textcolor{red}{14.44} & 171.29 & 0.27 \\
			\multicolumn{1}{c|}{\textbf{\textit{05}}} & 12.45 & 178.87 & \textcolor{blue}{0.20} & \multicolumn{1}{|c|}{\textbf{\textit{05}}} & \textcolor{red}{15.17} & \textcolor{blue}{102.66} & \textcolor{blue}{0.17} \\
			\multicolumn{1}{c|}{\textbf{\textit{06}}} & \textcolor{blue}{10.74} & \textcolor{red}{187.86} & \textcolor{red}{0.30} & \multicolumn{1}{|c|}{\textbf{\textit{06}}} & 14.12 & \textcolor{red}{182.55} & 0.25 \\
			\multicolumn{1}{c|}{\textbf{\textit{07}}} & \textcolor{red}{13.93} & 168.51 & \textcolor{red}{0.27} & \multicolumn{1}{|c|}{\textbf{\textit{07}}} & \textcolor{blue}{9.99}  & 173.61 & \textcolor{red}{0.35} \\
			\multicolumn{1}{c|}{\textbf{\textit{08}}} & 11.17 & \textcolor{red}{191.82} & 0.26 & \multicolumn{1}{|c|}{\textbf{\textit{08}}} & \textcolor{red}{15.16} & 135.16 & \textcolor{blue}{0.14} \\
			\multicolumn{1}{c|}{\textbf{\textit{09}}} & 12.49 & 164.82 & 0.25 & \multicolumn{1}{|c|}{\textbf{\textit{09}}} & 13.02 & \textcolor{red}{187.97} & \textcolor{red}{0.33} \\
			\multicolumn{1}{c|}{\textbf{\textit{10}}} & \textcolor{blue}{10.58} & 169.67 & 0.22 & \multicolumn{1}{|c|}{\textbf{\textit{10}}} & 12.97 & 168.71 & 0.30 \\
			\multicolumn{1}{c|}{\textbf{\textit{11}}} & \textcolor{red}{13.63} & 171.74 & \textcolor{blue}{0.18} & \multicolumn{1}{|c|}{\textbf{\textit{11}}} & 13.65 & 157.03 & 0.25 \\
			\multicolumn{1}{c|}{\textbf{\textit{12}}} & 12.89 & \textcolor{blue}{112.00} & 0.21 & \multicolumn{1}{|c|}{\textbf{\textit{12}}} & 12.28 & \textcolor{red}{180.91} & \textcolor{red}{0.31} \\
			\multicolumn{1}{c|}{\textbf{\textit{13}}} & \textcolor{blue}{10.59} & \textcolor{red}{186.17} & \textcolor{blue}{0.19} & \multicolumn{1}{|c|}{\textbf{\textit{13}}} & 13.77 & 128.10 & 0.27 \\
			\multicolumn{1}{c|}{\textbf{\textit{14}}} & \textcolor{red}{14.00} & 175.38 & 0.24 & \multicolumn{1}{|c|}{\textbf{\textit{14}}} & 12.67 & 169.37 & 0.27 \\
			\multicolumn{1}{c|}{\textbf{\textit{15}}} & 12.29 & 162.23 & 0.21 & \multicolumn{1}{|c|}{\textbf{\textit{15}}} & \textcolor{red}{14.64} & 150.07 & \textcolor{blue}{0.18} \\
			\multicolumn{1}{c|}{\textbf{\textit{16}}} & 11.24 & \textcolor{blue}{72.44}  & 0.25 & \multicolumn{1}{|c|}{\textbf{\textit{16}}} & 13.03 & 149.98 & 0.30 \\
			\multicolumn{1}{c|}{\textbf{\textit{17}}} & \textcolor{blue}{10.62} & \textcolor{blue}{159.53} & \textcolor{red}{0.35} & \multicolumn{1}{|c|}{\textbf{\textit{17}}} & 14.26 & \textcolor{blue}{84.85}  & 0.23 \\
			\multicolumn{1}{c|}{\textbf{\textit{18}}} & \textcolor{red}{14.09} & 173.38 & \textcolor{blue}{0.18} & \multicolumn{1}{|c|}{\textbf{\textit{18}}} & \textcolor{blue}{12.23} & \textcolor{blue}{74.75}  & 0.25 \\
			\multicolumn{1}{c|}{\textbf{\textit{19}}} & 12.33 & 176.98 & 0.26 & \multicolumn{1}{|c|}{\textbf{\textit{19}}} & 12.76 & 142.93 & 0.24 \\
			\multicolumn{1}{c|}{\textbf{\textit{20}}} & 12.70 & 180.76 & 0.25 & \multicolumn{1}{|c|}{\textbf{\textit{20}}} & 13.62 & 162.03 & 0.22 \\
			\bottomrule
		\end{tabular}
		\vspace{-1.1em}
	}
\end{table}

\begin{table}
	\renewcommand\arraystretch{1.1}
	\centering
	\scriptsize
	\caption{Quantitative Results of the USTC-TD 2023  Video Dataset. \\Note the Higher Scores are Represented in \textcolor{red}{Red}, and \\the Lower Scores are Represented in \textcolor{blue}{Blue}}
	\vspace{-0.6em}
	\label{tab:video_metrics}
	\setlength{\tabcolsep}{3.5mm}
	{
	\begin{tabular}{c|cc|cc}
		\toprule
		\multicolumn{5}{c}{\textbf{\textit{USTC-TD 2023}}} \\ 
		\midrule
		\multirow{2}{*}{Video Sequence} & \multicolumn{2}{c|}{Short Videos} & \multicolumn{2}{c}{Long Videos} \\
		\cmidrule(l){2-3} \cmidrule(lr){4-5}  
		& \textbf{\textit{SI}}       & \textbf{\textit{TI}}    & \textbf{\textit{SI}}       & \textbf{\textit{TI}}   \\
		\midrule
		\multicolumn{1}{c|}{\textbf{\textit{USTC-Badminton}}}      & 67.37  & 27.91 & 71.35 & 27.94\\
		\multicolumn{1}{c|}{\textbf{\textit{USTC-BasketballDrill}}}  & \textcolor{red}{130.31} & \color{red}{42.25} & 131.66 & \color{red}{44.19}\\
		\multicolumn{1}{c|}{\textbf{\textit{USTC-BasketballPass}}}   & 94.63  & \color{red}{46.92} & 95.74 & \color{red}{46.89}\\
		\multicolumn{1}{c|}{\textbf{\textit{USTC-BicycleDriving}}}   & \textcolor{blue}{38.66}  & \color{red}{52.70} & 96.09 & \color{red}{52.73}\\
		\multicolumn{1}{c|}{\textbf{\textit{USTC-Dancing}}}          & 74.34  & \color{blue}{19.61} & 7.22 & \color{blue}{26.16} \\
		\multicolumn{1}{c|}{\textbf{\textit{USTC-FourPeople}}}       & 45.49  & \color{blue}{9.54} & 45.77 & \color{blue}{9.59} \\
		\multicolumn{1}{c|}{\textbf{\textit{USTC-ParkWalking}}}      & 72.48  & 40.22  & 119.07 & 40.40\\
		\multicolumn{1}{c|}{\textbf{\textit{USTC-Running}}}          & 86.52  & 34.84 & 89.02 & 34.97\\
		\multicolumn{1}{c|}{\textbf{\textit{USTC-ShakingHands}}}     & \textcolor{red}{100.67} & \color{red}{44.11} & 100.74 & \color{red}{48.37}\\
		\multicolumn{1}{c|}{\textbf{\textit{USTC-Snooker}}}          & \textcolor{blue}{41.62}  & 29.63 & 50.01 & 35.85 \\
		\bottomrule
	\end{tabular}
		\vspace{-1.9em}
	}
\end{table}

Specifically, for the evaluation of USTC-TD image dataset, the scores of SI, LI, CF cooperate to evaluate its spatial, colorfulness, and lightness diversity. Compared to the other image datasets\cite{Kodak, clic, asuni2014testimages}, it exhibits SI scores ranging from 9 to 16, and is distinguished from other image datasets, as shown in Fig.\,\ref{fig:Image_CF_SI}\,. The proposed image dataset incorporates more spatial diversity within the wide range of colorfulness diversity. In Fig.\,\ref{fig:light_SI_LI}\,, the proposed image dataset also shows a wide coverage of LI scores, ranging from 0.10 to 0.40, which aligns with the range of other image datasets and demonstrates that the proposed image dataset also exhibits excellent generalization of lightness diversity. For the evaluation of USTC-TD video dataset (short/long setting), the scores of TI and SI cooperate to evaluate its spatial and temporal diversity, as shown in Fig.\,\ref{fig:video_SI_TI}\,. Compared to the other video datasets\cite{bossen2013common, boyce2018jvet, wang2016mcl, mercat2020uvg}, the proposed video dataset incorporates more temporal diversity within the excellent generalization of spatial diversity. For the short setting, it exhibits a wide range of temporal variation, ranging from 5 to 55. It compensates for the absence of the 40 to 55 (higher) range in the temporal variation of other video datasets, which enables an excellent coverage of temporal diversity with the collaboration of other datasets. For the long setting, it further extends the coverage of temporal variation, enabling a more robust and comprehensive assessment of video compression schemes. The above analysis results and related codes are open-sourced.

\subsubsection{Driving Factors Behind the Benefit of USTC-TD}
Beyond the above feature analysis, we present the detailed variety distribution of these different features of USTC-TD image/video datasets, as shown in Table~\ref{tab:image_metrics}\, and \ref{tab:video_metrics}\,, the detailed scores of different evaluative features are represented with the annotation of different colors. Here we deeply discuss the driving factors behind the benefit of USTC-TD image and video datasets.

\begin{table*}
	\renewcommand\arraystretch{1.6}
	\centering
	\vspace{-1.5em}
	\scriptsize
	\caption{Three Recommended Presets of the Existing Image/Video Datasets \cite{Kodak, asuni2014testimages, clic, mercat2020uvg, wang2016mcl, bossen2013common, boyce2018jvet} and the Proposed USTC-TD  \\ for the Practical Evaluation of Image/Video Compression Schemes}
	\vspace{-0.6em}
	\label{tab:dataset_preset}
	\setlength{\tabcolsep}{2mm}
	{
		\begin{tabular}{c|c|c|c|c}
			\hline
			\textbf{Preset}    & \textbf{Image Dataset}                 & \textbf{Video Dataset}                    & \textbf{Characteristic} & \textbf{Purpose} \\ \hline
			\textit{Challenging} & \textit{USTC-TD}                       & \textit{USTC-TD}                          & Challenging Content                & Developing Advanced Coding Algorithms          \\ \hline
			\textit{Desirable}  
			& \textit{USTC-TD}, 
			\textit{Tecnick}\cite{asuni2014testimages}              & \textit{USTC-TD}, 
			\textit{MCL-JCV}\cite{wang2016mcl}, \textit{HEVC CTC}\cite{bossen2013common}            & Wider Coverage                  & Obtaining Reliable Estimation of Coding Performance \\ \hline
			\textit{Ideal}      
			& \begin{tabular}[c]{@{}c@{}}\textit{USTC-TD}, 
			\textit{CLIC}\cite{clic},\vspace{-0.4em}\\
			\textit{Tecnick}\cite{asuni2014testimages}, 
			\textit{Kodak}\cite{Kodak}\end{tabular} 
			& \begin{tabular}[c]{@{}c@{}}\textit{USTC-TD}, 
			\textit{UVG}\cite{mercat2020uvg}, 
			\textit{MCL-JCV}\cite{wang2016mcl}, \vspace{-0.4em}\\
			\textit{HEVC CTC}\cite{bossen2013common}, 
			\textit{VVC CTC}\cite{boyce2018jvet}\end{tabular} 
			 & Sufficient Coverage           & \begin{tabular}[c]{@{}c@{}}Evaluating Algorithm Robustness \vspace{-0.4em} \\and Generalization Ability \end{tabular}\\ \hline
		\end{tabular}
		\vspace{-1.2em}
	}
\end{table*}

For the benefit of USTC-TD image dataset, it compensates for the absence of the higher range of spatial variation within the generalized colorfulness, lightness variation and adds the mixture content diversity, which distinctly differentiates it from existing image datasets\cite{Kodak, asuni2014testimages, clic}. Compared with the previous image datasets, the driving factors of the benefit mainly come from two aspects: the specific design of various content factors (environmental/imaging-related factors), and the mixture of content diversity (spatial, lightness, colorfulness diversity), which contribute to maximizing the coverage. In detail, first, the samples with higher spatial diversity exhibit the characteristics of complex content factors. For example, the samples with SI scores higher than 13 of Fig.\,\ref{fig:Image_CF_SI}\, have the above aspect, such as the \textit{USTC-TD-2022-01, 14, 18}, \textit{USTC-TD-2023-04, 05, 08}. According to the Table~\ref{tab:ID2022}\, and \ref{tab:ID2023}\,, these samples own the specific design of complex environmental or imaging factors, such as the \textit{USTC-TD-2022-01, 18, USTC-TD-2023-04} with the typical texture (geometric), or \textit{USTC-TD-2022-14/USTC-TD-2023-05} with extreme illumination (underexposure/overexposure), or \textit{USTC-TD-2023-08} with the special captured view (overhead level). Second, the samples with mixture diversity enable the outstanding coverage of different features, such as the \textit{USTC-TD-2022-07, 11, USTC-TD-2023-05, 15}. As shown in Table~\ref{tab:image_metrics}\,, these samples own the mixture of different kinds of content diversity, such as the \textit{USTC-TD-2022-07} with higher spatial and lightness diversity (higher SI, LI scores), the \textit{USTC-TD-2023-05} with various diversity (higher SI, Lower CF, LI scores). The mutual promotion of different components of content factors has further contributed to the maximization of coverage. 

For the benefit of USTC-TD video dataset, it compensates for the absence of the higher range of temporal variation within the generalized spatial variation, which distinctly differentiates it from existing video datasets\cite{bossen2013common, boyce2018jvet, wang2016mcl, mercat2020uvg}\,. Compared with the previous video datasets, the driving factors of the benefit directly come from the  augmentation and extension of motion diversity. For example, according to Table~\ref{tab:VD2023} and \ref{tab:video_metrics}\,, the samples with TI scores higher than 40 of Fig.\,\ref{fig:video_SI_TI}\, exhibit the characteristics of diverse motion factors, such as \textit{USTC-BicycleDriving} with high-speed moving objects and scene change, \textit{USTC-ShakingHands} with precise limb motion, \textit{USTC-BaskerballDrill} with abundant athletes' motion and rich lens moving. Compared with the characteristics of existing datasets, the above different kinds of motion factors further compensate for the limitation of temporal correlation, and contribute to the robust evaluation and the maximization of temporal relation-related property coverage. 

\subsubsection{Discussion of Practical Utilization and Application of the Existing Compression Datasets and USTC-TD}
Beyond the feature and benefit analysis of USTC-TD, it has been verified that the USTC-TD compensates for the shortcomings of existing image/video datasets, and its collaboration with existing datasets can achieve wider coverage of different image/video features. Although the collaboration of these datasets ensures the robust evaluation of advanced compression schemes, the evaluation of all image/video datasets will cause heavy time/computational complexity in practical use. Here we discuss their desirable collaboration for future compression research, and further put forward the different recommended presets of USTC-TD and existing image/video datasets with the consideration of practical utilization. 

To achieve an efficient evaluation for future research, we discuss the desirable collaboration of existing image/video datasets and USTC-TD from two perspectives: \textit{challenge} and \textit{coverage}, and present three presets of these image and video datasets to support them with the consideration of different uses, including \textit{challenging}, \textit{desirable}, and \textit{ideal} presets, as shown in Table~\ref{tab:dataset_preset}\,. For the challenge, in the development of compression standardization, the coding efficiency on existing test content has gradually reached a bottleneck, such as the performance of hand-crafted intra and inter prediction, making it increasingly difficult to reflect shortcomings and guide future improvements. Thus, incorporating challenging content is essential, as it serves as a continuous driving force for the improvement of next-generation coding standards. For \textit{challenging} preset, typical challenging samples are considered for developing advanced coding algorithms. Correspondingly, USTC-TD samples with the highest spatial/temporal diversity (scores) are chosen for evaluation. For coverage, building upon the \textit{challenging} preset, a broader coverage of diverse content is desirable for the consistent evaluation of algorithm performance. By incorporating a wider range of different kinds of content features, test datasets can effectively evaluate the codec's performance in various scenarios, ensuring its reliability in practical use. For \textit{desirable} preset, the samples with the highest coverage of different kinds of content diversity (spatial, colorfulness, lightness for image samples, temporal, spatial for video samples) are considered for obtaining reliable estimation of coding performance, such as the collaboration of \textit{USTC-TD} image dataset and \textit{Tecnick} \cite{asuni2014testimages}, \textit{USTC-TD} video dataset, \textit{MCL-JCV}\cite{wang2016mcl}, and \textit{HEVC CTC}\cite{bossen2013common}. These collaborations cover the majority of samples with different kinds of content diversity, with an appropriate number of representative samples supporting comprehensive evaluation in practical use. For the \textit{ideal} preset, given sufficient time/resources, all datasets are ideally tested to ensure a comprehensive evaluation of algorithm robustness and generalization, supporting its use in rigorous benchmarking and evaluation.

\vspace{-1.2em}
\section{Experiments}
\vspace{-0.4em}
In this section, first, we present the experimental configurations employed for the evaluation of compression schemes. Second, we evaluate the classic standardized compression schemes and recent advanced image/video compression schemes on the proposed dataset under different metrics, and benchmark their performance on our proposed dataset. Third, we analyze the benchmarked performance, and further point out some limitations and inspirations among these advanced image/video schemes to shed light on future research.

\vspace{-1em}
\subsection{Experimental Settings}
\vspace{-0.1em}
In this subsection, first, we present the experimental settings of the evaluative compression schemes, including the selection of advanced image/video compression schemes and the training/testing configurations of these compression schemes. Second, we introduce the evaluative quality metrics for these schemes on our proposed dataset.

\subsubsection{{Selection of Evaluative Compression Schemes}}
For advanced image compression schemes, we select classic standardized schemes and advanced learned schemes. For traditional codecs, we select \textit{BPG}\,\footnote{Available online at \,\url{https://bellard.org/bpg}.}, and \textit{H.266/VVC} \cite{bross2021overview}. For learned image compression schemes, we select the \textit{Factorized Model}\cite{balle2018variational},
\textit{Hyperprior Model}\cite{balle2018variational},
\textit{Autoregressive Model}\cite{minnen2018joint},
\textit{Cheng2020}\cite{cheng2020learned}, \textit{iWave++}\cite{ma2020end}, \textit{ELIC}\cite{he2022elic}, and \textit{MLIC++}\cite{jiang2023mlic}. For learned compression standardized schemes, we select the high-profile model of \textit{IEEE 1857.11}\,\footnote{Available online at \,\url{https://sagroups.ieee.org/fvc/}.} (\textit{iWave++}).

For video compression schemes, we also select classic standardized schemes and advanced learned schemes. For standardized codecs, we select the \textit{H.265/HEVC}\cite{sullivan2012overview}, \textit{H.266/VVC}\cite{bross2021overview}, \textit{AV1}\cite{han2021technical}, and \textit{AV2} \cite{zhao2021study}. For learned video compression schemes, we select the 
\textit{DVC\_Pro}\cite{lu2019dvc, lu2020end}, \textit{CANF-VC}\cite{ho2022canf}, \textit{DCVC}\cite{li2021deep},
\textit{TCM-VC}\cite{sheng2022temporal},
\textit{DCVC-HEM}\cite{li2022hybrid},
\textit{OOFE}\cite{tang2024offline},
\textit{SDD}\cite{sheng2024spatial},
\textit{VNVC}\cite{sheng2024vnvc},
\textit{DCVC-DC}\cite{li2023neural}, and
\textit{DCVC-FM}\cite{li2024neural}. The detailed introduction and  test instructions of these methods are mentioned in Section I of supplementary material.

\subsubsection{Testing Configurations of Evaluative Traditional Image Compression Schemes}
For testing, the officially released BPG software, VTM-17.0 (\textit{H.266/VVC} reference software) are chosen. For \textit{BPG}, the default configuration is used, and the internal color space is set to YUV420/444 for the testing of \textit{BPG} and \textit{BPG444}. For VTM-17.0, the \textit{encoder\_intra\_vtm} configuration is used, and the internal color space is set to YUV444. For the different source formats of these testing datasets (USTC-TD and \cite{Kodak, asuni2014testimages, clic}), we convert them to the YUV444 color space for the input of VTM-17.0 by using the default \textit{ffmpeg} tool (BT.601 conversion standard).

\subsubsection{Testing Configurations of Evaluative Traditional Video Compression Schemes}
For testing, the officially released HM-16.20 (\textit{H.265/HEVC} reference software), VTM-13.2 (\textit{H.266/VVC} reference software), AV1-3.11.0, and AV2-7.0.0 are chosen. For the setting of HM-16.20, the \textit{encoder\_lowdelay\_main\_rext} configuration is used. For the setting of VTM-13.2, the \textit{encoder\_lowdelay\_vtm} configuration is used. For the different source formats of these testing datasets (USTC-TD and \cite{mercat2020uvg, wang2016mcl, bossen2013common, boyce2018jvet}), we convert them to the YUV444 color space as the input of the above traditional codecs by using the default \textit{ffmpeg} tool, the conversion step is aligned to the setting mentioned in \textit{DCVC-DC}\cite{li2023neural}.

\begin{figure*}[!t]
	\vspace{-1em}
	\captionsetup[subfigure]{labelformat=empty}
	\centering
	\includegraphics[width=1.6in]{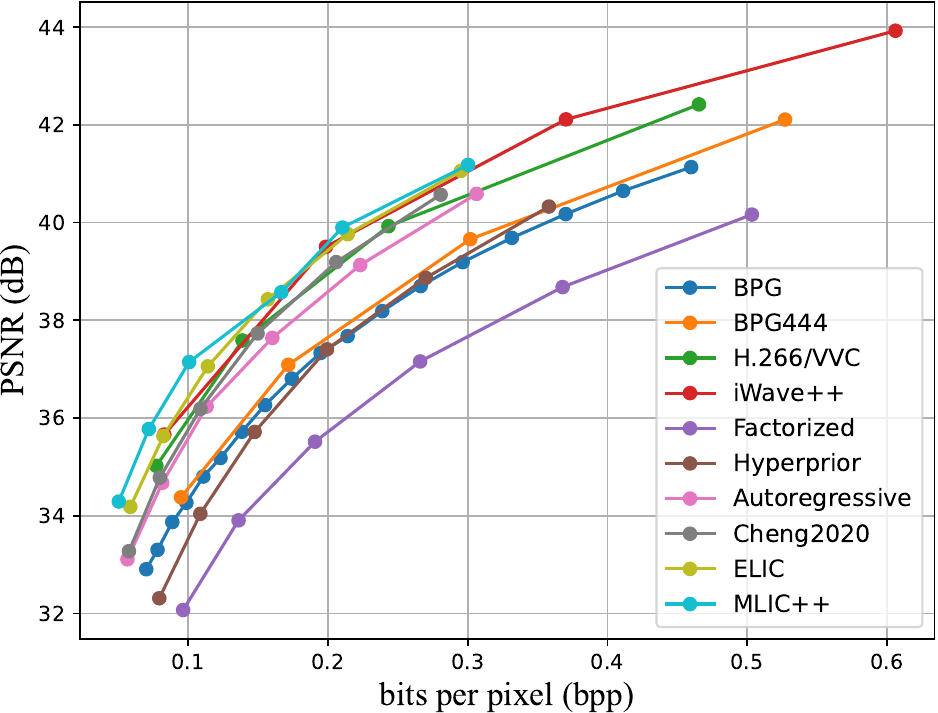}
	\includegraphics[width=1.64in]{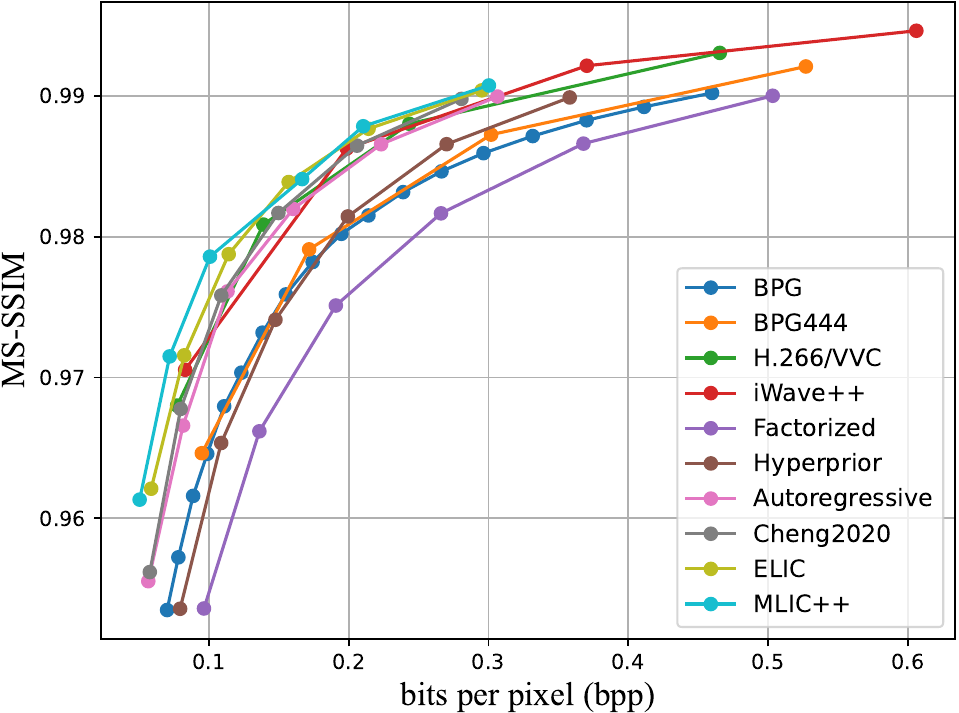}
	\includegraphics[width=1.6in]{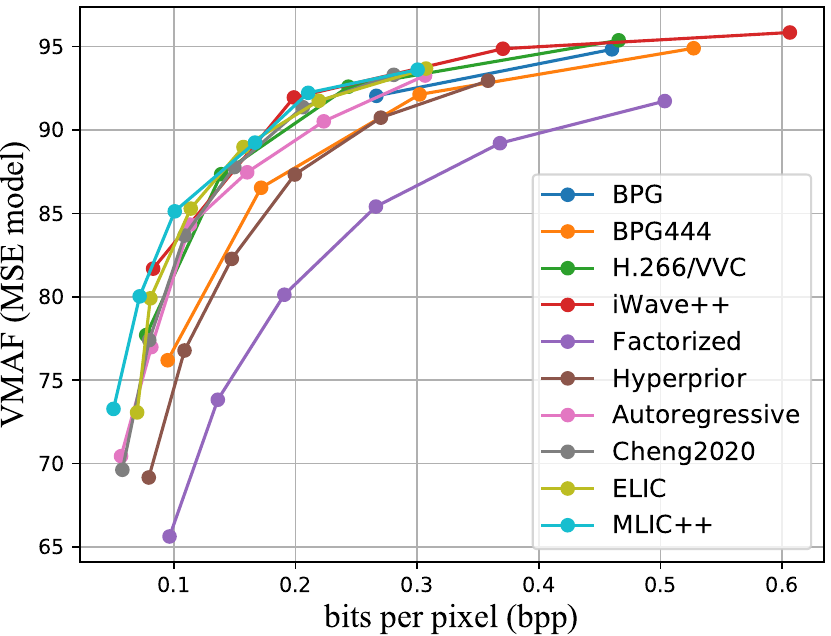}
	\hspace{0.7mm}\includegraphics[width=1.61in]{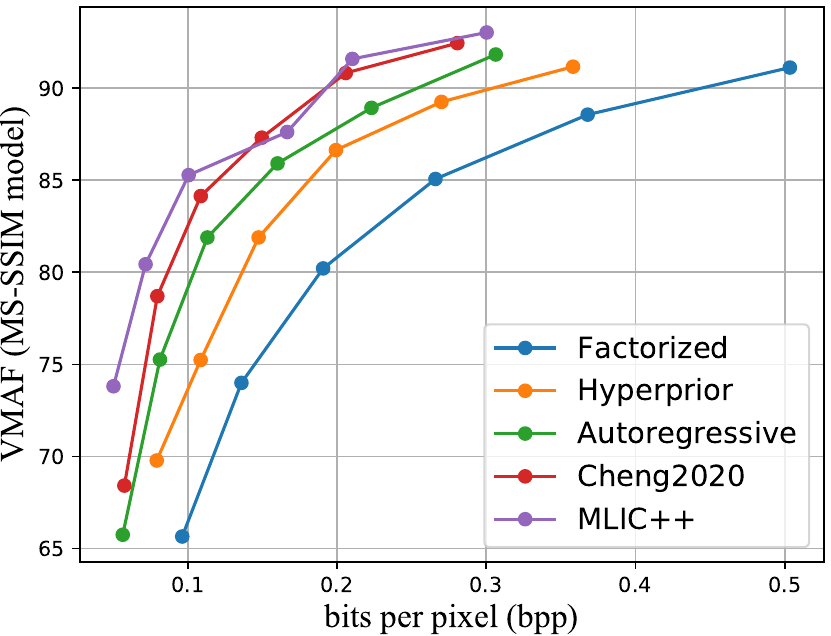}
	\vspace{-0.7em}
	\caption{Overall rate-distortion (RD) curves of advanced image compression schemes on different metrics. From left to right, the results are evaluated by \textit{PSNR}, \textit{MS-SSIM}, \textit{VMAF (MSE model)}, and \textit{VMAF (MS-SSIM model)} metrics on USTC-TD image dataset 2022 and 2023. \label{fig:img_results_avg}}
	\vspace{-0.3em}
\end{figure*}

\begin{figure*}[!t]
	\centering
	\includegraphics[width=1.6in]{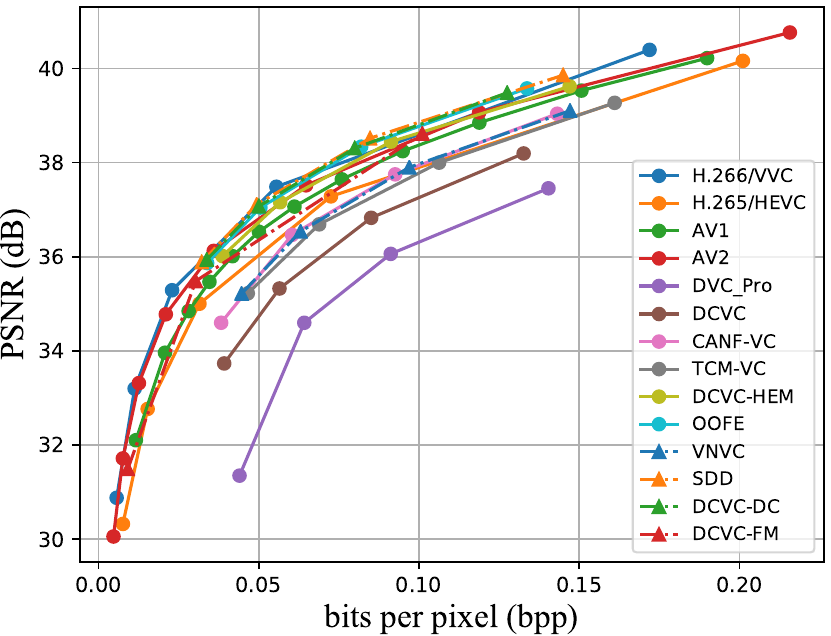}
	\includegraphics[width=1.64in]{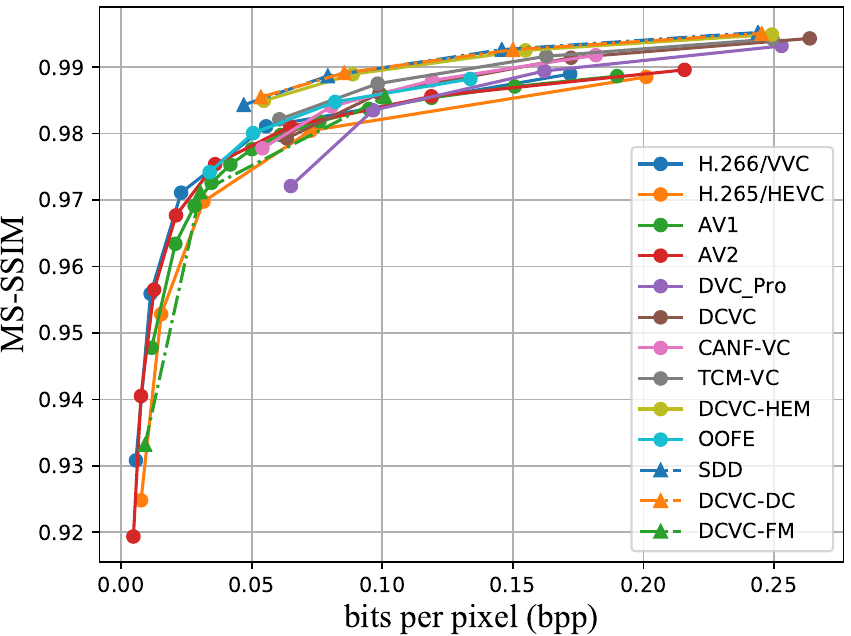}
	\includegraphics[width=1.61in]{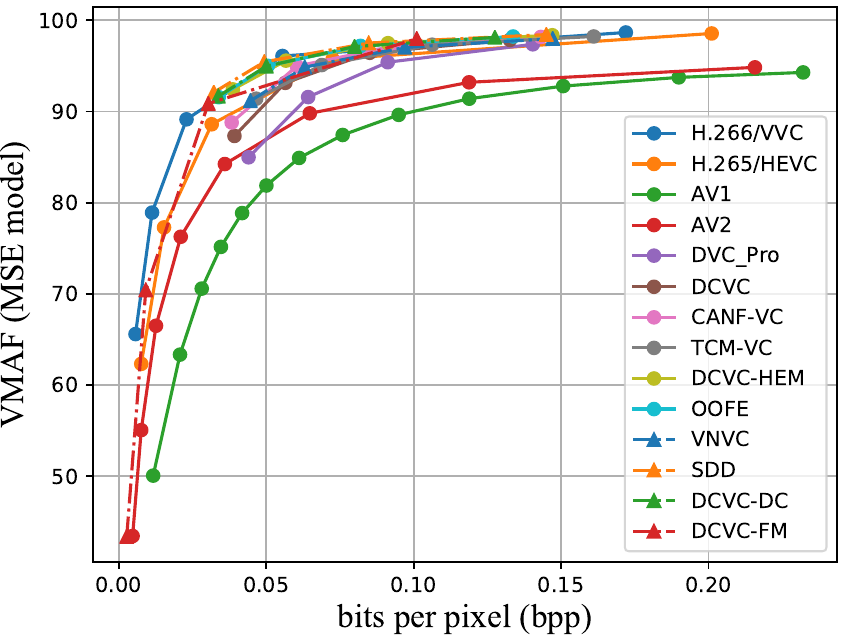}
	\includegraphics[width=1.6in]{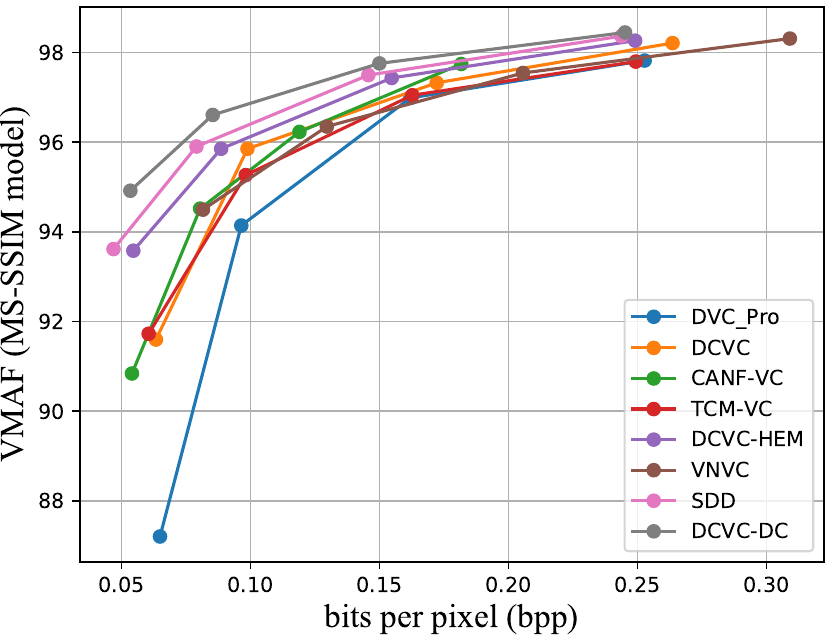}
	
	\footnotesize
	\makebox[\textwidth]{(a) Short Setting, Intra Period = 32}\vspace{-0.6em}
	
	\rule{\textwidth}{0.5pt}
	
	\vspace{0.7em}
	\includegraphics[width=1.6in]{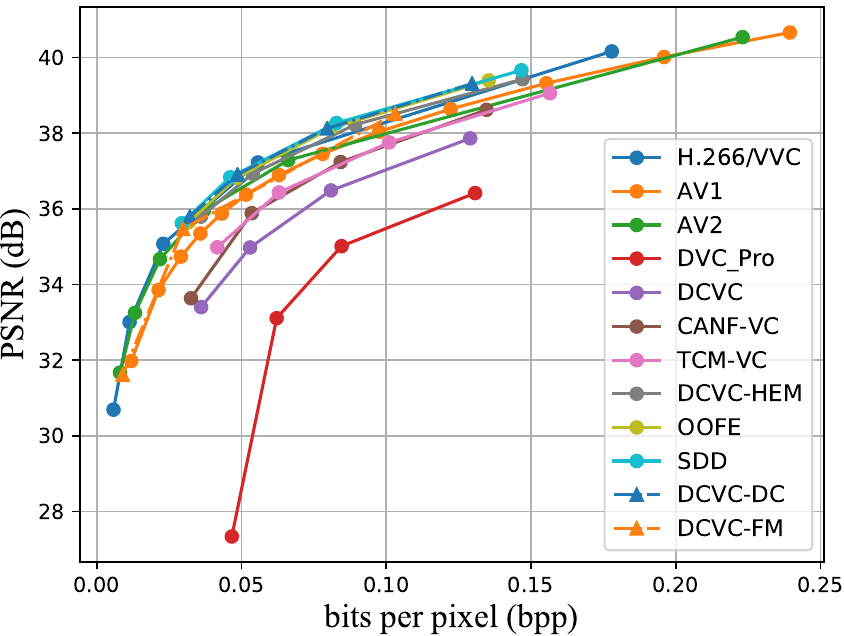}
	\includegraphics[width=1.61in]{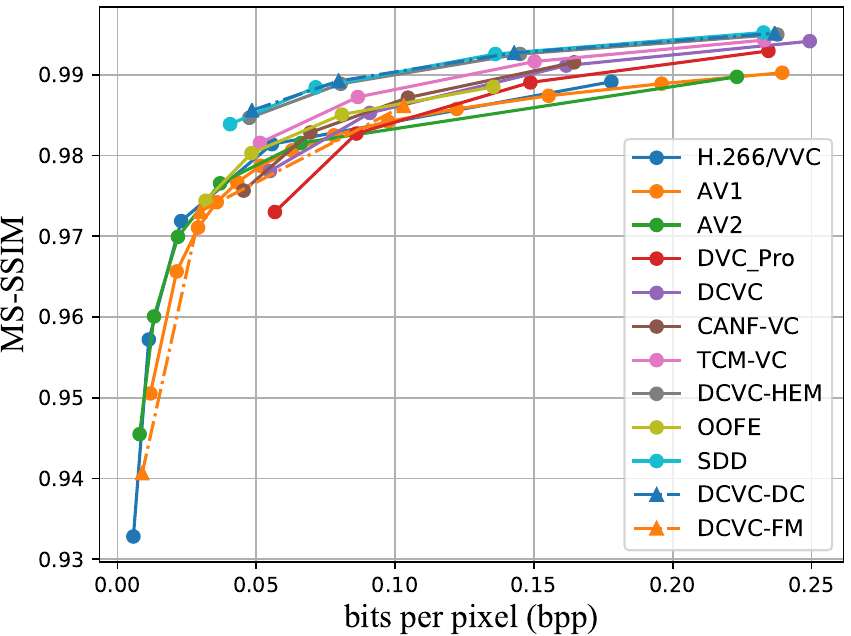}
	\hspace{0.3em}\includegraphics[width=1.61in]{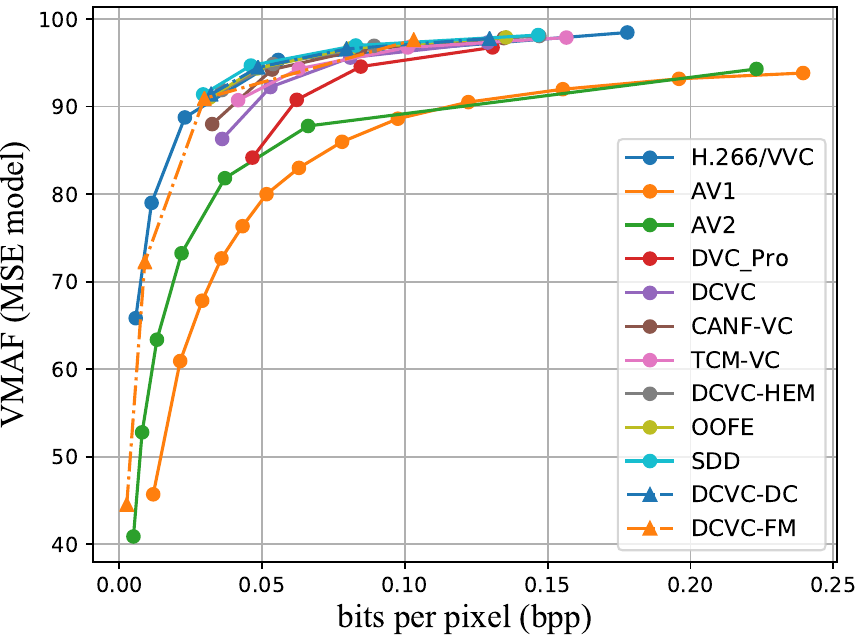}
	\includegraphics[width=1.6in]{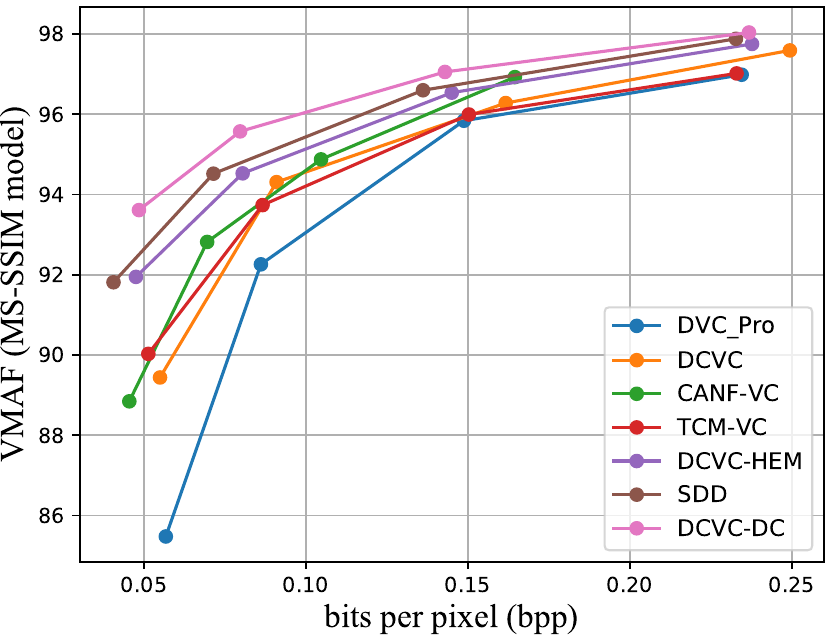}
	
	\footnotesize
	\makebox[\textwidth]{(b) Long Setting, Intra Period = 32}\vspace{-0.5em}
	
	\rule{\textwidth}{0.5pt}
	
	\vspace{0.9em}
	\includegraphics[width=1.6in]{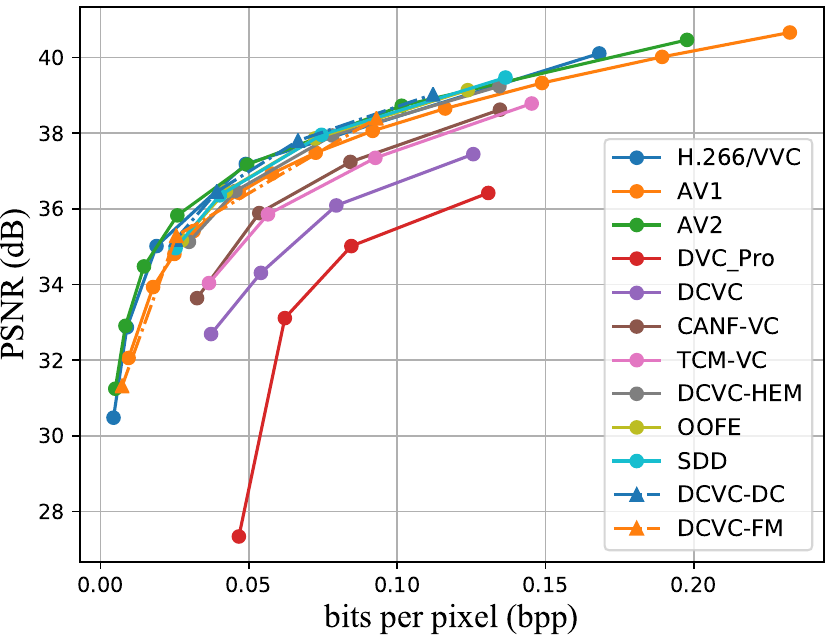}
	\includegraphics[width=1.63in]{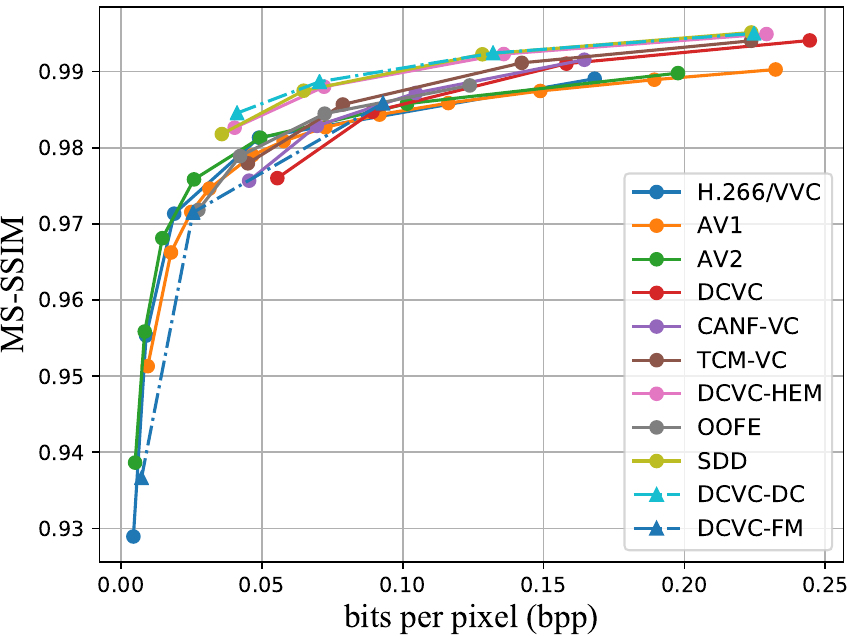}
	\hspace{0.3em}\includegraphics[width=1.6in]{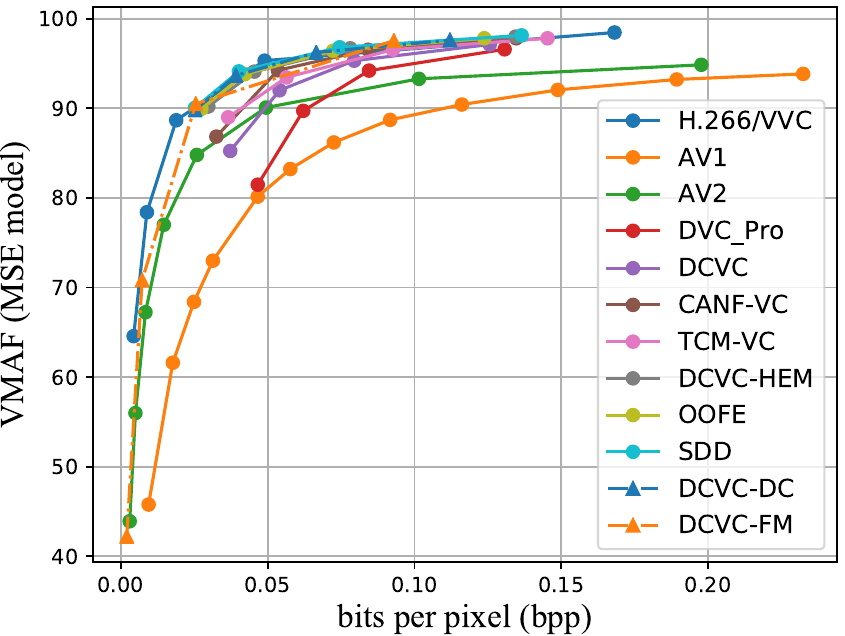}
	\hspace{0.2em}\includegraphics[width=1.6in]{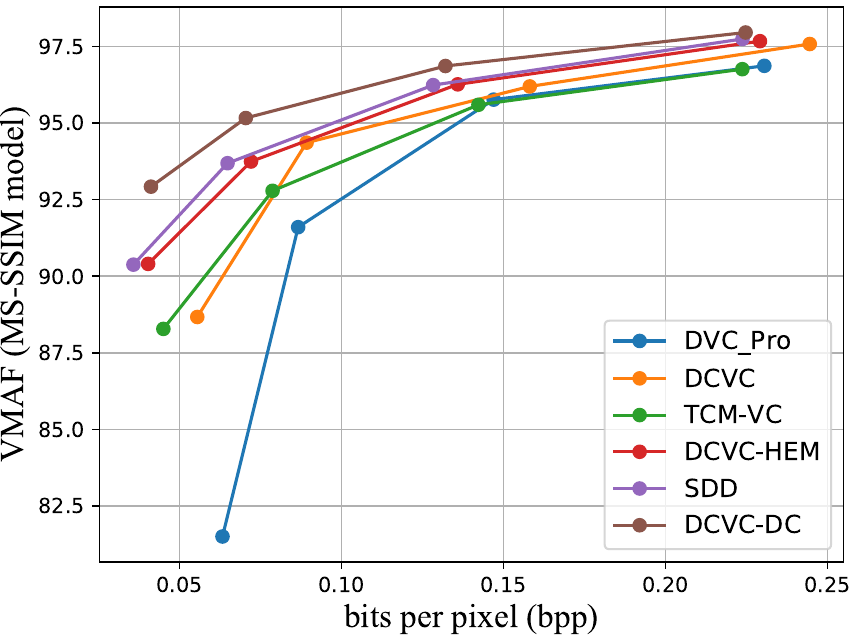}
	
	\footnotesize
	\makebox[\textwidth]{(c) Long Setting, Intra Period = -1}
	
	\vspace{-0.8em}
	
	\caption{Overall rate-distortion (RD) curves of advanced video compression schemes on different metrics. From left to right, the results are evaluated by \textit{PSNR}, \textit{MS-SSIM}, \textit{VMAF (MSE model)}, and \textit{VMAF (MS-SSIM model)} metrics on different settings of USTC-TD video dataset.}
	\vspace{-1.8em}
	\label{fig:video_results_avg}
\end{figure*}

\subsubsection{Training and Testing Configurations of Evaluative Learned Image Compression Schemes}
For training, these learned image compression models are optimized by mean squared error (\textit{MSE}) or multi-scale structural similarity index measure (\textit{MS-SSIM}), and the \textit{Flicker2W} \cite{liu2020unified} dataset is used as the training dataset. These models are optimized by using the Adam Optimizer \cite{kingma2014adam}, with a batch size of 8 and a patch size of $256\times 256$. They are optimized for around 1.2 million iterations, starting with an initial learning rate of $10^{-4}$. The learning rate is reduced to $10^{-5}$ after 400 epochs and further down to $10^{-6}$ after 30 epochs. The setting of $\lambda$ is set to $\{0.001,0.004,0.024,0.080,0.200\}$ for iWave++, and $\{0.0018,0.0035,0.0067,0.0130,0.0250,0.0483\}$ for other schemes. For testing, the officially released model of \textit{MLIC++}/\textit{iWave++}, the reproduced model of \textit{ELIC} are used. The model of other schemes is provided by \textit{CompressAI} \cite{begaint2020compressai}. 

\subsubsection{Training and Testing Configurations of Evaluative Learned Video Compression Schemes}
For training, these learned video compression models are mainly optimized by \textit{MSE} or \textit{MS-SSIM}, and the \textit{Viemo-90k}\cite{xue2019video} is used as the training dataset. For the testing of USTC-TD, the officially released models of these schemes are used. For the testing of the variable-rate model, the bitrate points are aligned to that of the traditional codec. For the testing of other models, we directly use the released model. For the different formats of these testing datasets, we convert them to the RGB color space as the input of different learned video codecs by using the default \textit{ffmpeg} tool (BT.601), the conversion step is aligned to the setting mentioned in \textit{DCVC-DC}\cite{li2023neural}.

\subsubsection{Testing Configurations of Subjective Quality Evaluation of Image and Video Compression Schemes}
For the testing of traditional compression model, we directly use their officially released standardized model, these hand-crafted models are designed and optimized for \textit{PSNR} metric. For the testing of learned compression model, we use the officially released \textit{MS-SSIM} model of the advanced learned image and video compression schemes, these trained models are optimized for \textit{MS-SSIM} metric. Some image compression schemes \cite{ma2020end,he2022elic} without the open-sourced \textit{MS-SSIM} model are skipped. For the testing of video compression model, the long USTC-TD video dataset is used. To efficiently conduct the subjective test, we select the compressed images/video results at the intermediate bitrate point among all test bitrate points. For the evaluation of image compression schemes, since most learned approaches do not support the variable bitrate models, we align the target bitrate point of all approaches to approximately 0.12 bits per pixel. For the evaluation of video compression schemes, we align the target bitrate point of all approaches to approximately 0.06 bits per pixel.

\subsubsection{Evaluative Metrics} 
For the objective quality evaluation, \textit{PSNR}, \textit{MS-SSIM}\cite{wang2003multiscale}, and \textit{VMAF}\cite{liu2013visual} are used to measure the quality of the coded frames in comparison to the original frames. Bits per pixel (bpp) is used to measure the number of bits for encoding each pixel in each image/video frame. The Bjontegaard Delta bitrate (BD-rate)\cite{bjontegaard2001calculation} is used to compare the performance of different compression schemes, where negative numbers indicate bitrate saving and positive numbers indicate bitrate increasing. For the evaluation of image/video compression schemes, \textit{PSNR} and \textit{MS-SSIM} are calculated and compared in RGB color space, \textit{VMAF} is calculated and compared in YUV color space. The conversion process of different color spaces is performed by using the default \textit{ffmpeg} tool (BT.601). For \textit{PSNR} and \textit{MS-SSIM} tests, the \textit{MSE} model of these schemes is used for \textit{PSNR}, while the \textit{MS-SSIM} model of these schemes is used for \textit{MS-SSIM}. For \textit{VMAF} test, the \textit{MSE} and \textit{MS-SSIM} models are all used for \textit{VMAF} test (\textit{VMAF MSE/MS-SSIM model} for short). Notably, for some test schemes without the \textit{MS-SSIM} model, their \textit{MSE} model is used for the \textit{MS-SSIM} test, and the \textit{VMAF (MS-SSIM model)} test is skipped.

For the subjective quality evaluation, mean opinion score (\textit{MOS}) \cite{rec2006p, international1996methods, itu2017vocabulary, series2012methodology} is used to quantify the perceptual quality of the coded frames based on human evaluations, which is calculated by collecting ratings from multiple participants who assess the quality of images/videos on a pre-defined five-point scale (5$\sim$Excellent, 4$\sim$Good, 3$\sim$Fair, 2$\sim$Poor, 1$\sim$Bad). The final \textit{MOS} score is computed as the average of ratings given by all participants across test samples, the formula is as follows:
\begin{equation}
	MOS = \frac{1}{M} \sum_{i=1}^{M} \left( \frac{1}{N} \sum_{j=1}^{N} S_{i,j} \right)
\end{equation}
where \( M \) is the total number of test samples, \( N \) is the number of participants, and \( S_{i,j} \) represents the rating given by the \( j \)-th participant for the \( i \)-th sample. In our setting, 50 participants are selected for the \textit{MOS} test, with testers from USTC and online sources, ensuring diversity in user experience and perception. 

\vspace{-0.5em}
\subsection{Experimental Results}
In this subsection, we establish the baselines and benchmark the performance of advanced image/video compression schemes on USTC-TD image/video datasets, and further analyze their performance.

\subsubsection{Objective Quality Evaluation and Analysis of Advanced Image Compression Schemes on USTC-TD}
\begin{figure*}[!t]
	\centering
	\vspace{-1em}
	
	\begin{minipage}{0.22\textwidth}
		\centering
		\includegraphics[width=1.5in]{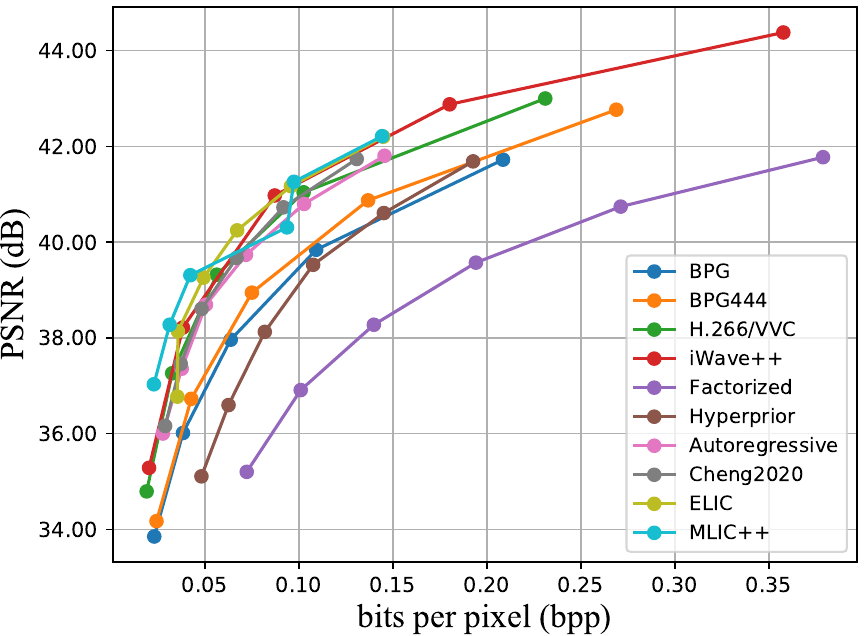} \\
		\scriptsize \hspace{0.3em} (a) USTC-2022-12
	\end{minipage}
	\begin{minipage}{0.22\textwidth}
		\centering
		\includegraphics[width=1.52in]{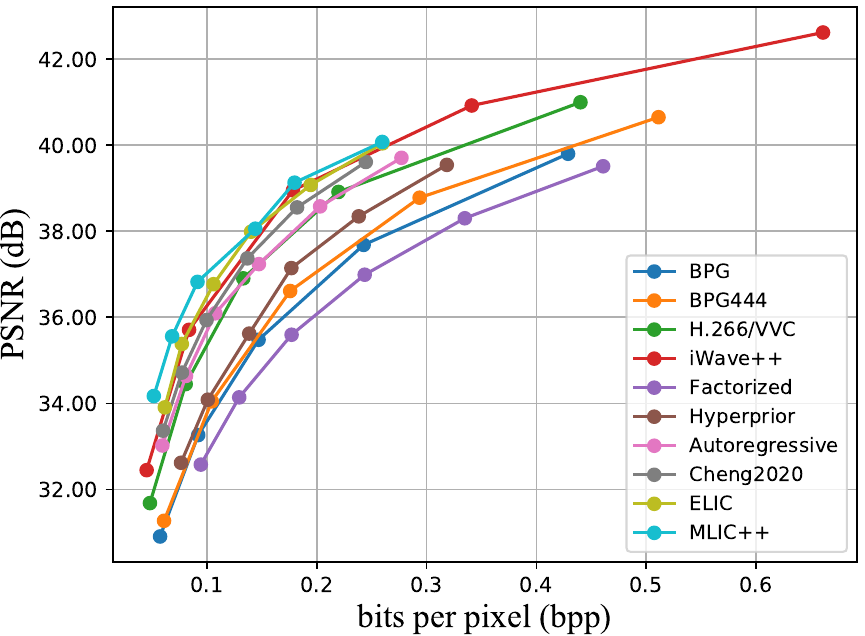} \\
		\scriptsize \hspace{0.8em}(b) USTC-2023-15
	\end{minipage}
	\begin{minipage}{0.22\textwidth}
		\centering
		\includegraphics[width=1.5in]{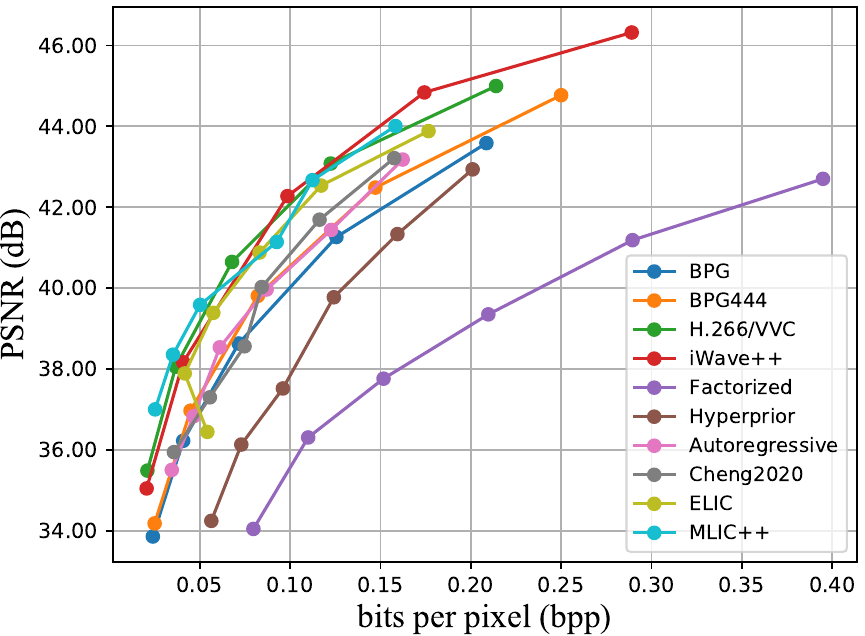} \\
		\scriptsize \hspace{1em}(c) USTC-2022-17
	\end{minipage}
	\begin{minipage}{0.22\textwidth}
		\centering
		\includegraphics[width=1.5in]{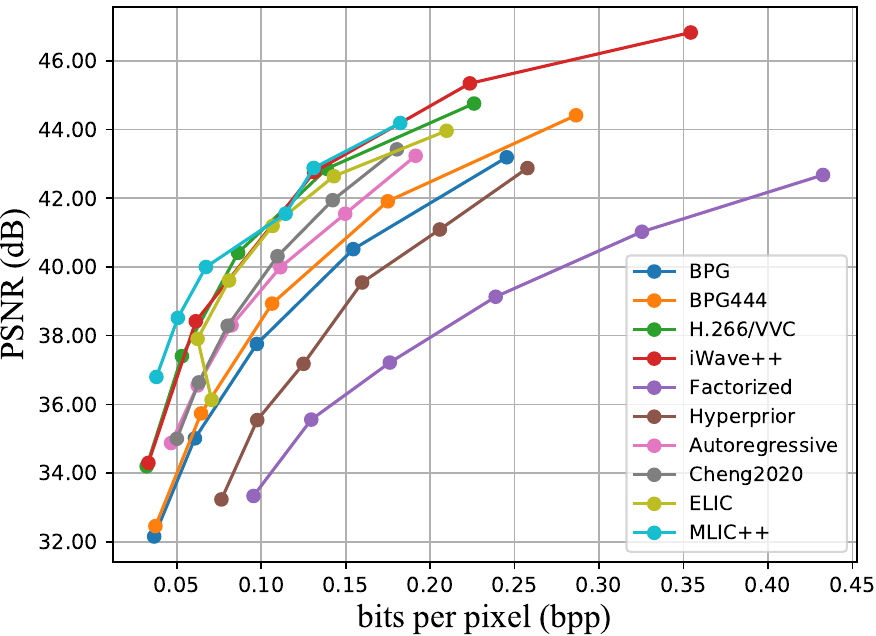} \\
		\scriptsize \hspace{1.1em}(d) USTC-2023-07
	\end{minipage}
	
	\vspace{-0.3em}
	\caption{Specific rate-distortion (RD) curves of advanced image compression schemes on partial evaluative images under \textit{PSNR} metric.}
	\label{fig:drd-image}
	
	\centering
	\vspace{0.7em}
	
	\begin{minipage}{0.22\textwidth}
		\centering
		\includegraphics[width=1.5in]{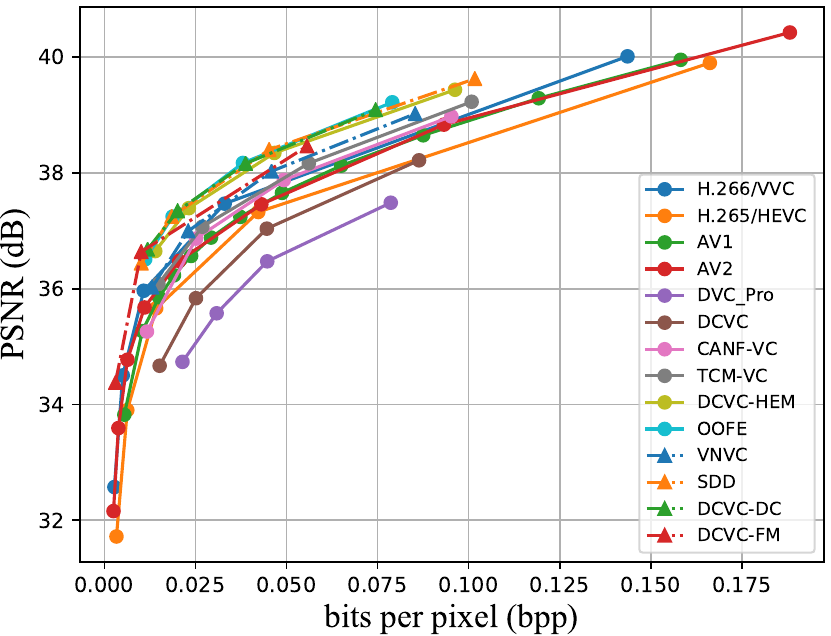} \\
		\scriptsize \hspace{0.3em}(a) USTC-FourPeople
	\end{minipage}
	\begin{minipage}{0.22\textwidth}
		\centering
		\includegraphics[width=1.52in]{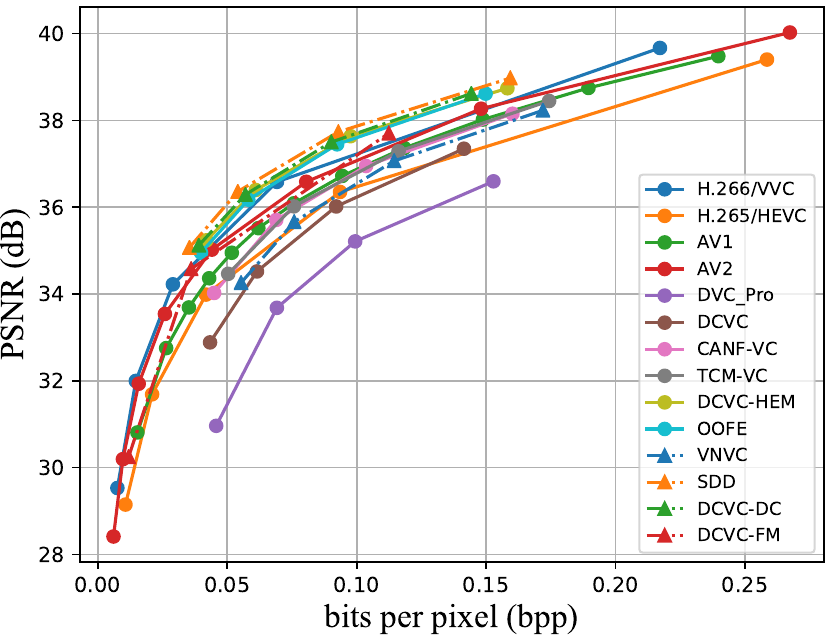} \\
		\scriptsize \hspace{0.8em}(b) USTC-BasketballPass
	\end{minipage}
	\begin{minipage}{0.22\textwidth}
		\centering
		\includegraphics[width=1.5in]{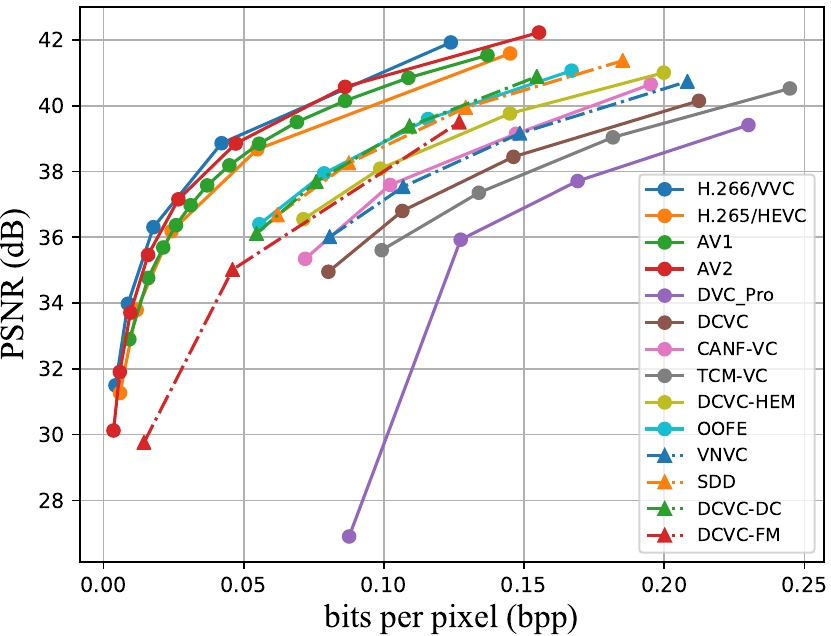} \\
		\scriptsize \hspace{1em}(c) USTC-BicycleDriving
	\end{minipage}
	\begin{minipage}{0.22\textwidth}
		\centering
		\includegraphics[width=1.5in]{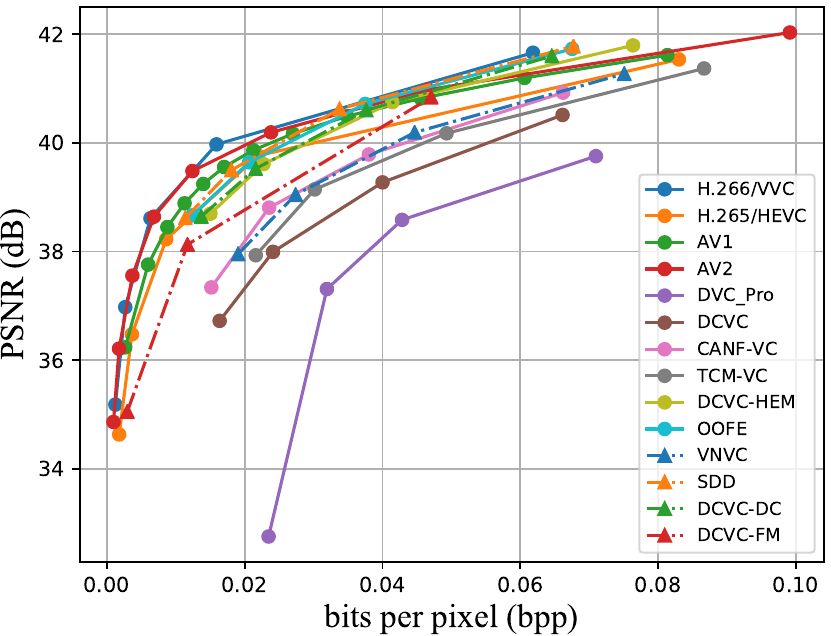} \\
		\scriptsize \hspace{1.1em}(d) USTC-Snooker
	\end{minipage}
	
	\vspace{-0.1em}
	\caption{Specific rate-distortion (RD) curves of advanced video compression schemes on partial evaluative short videos under \textit{PSNR} metric.} 
	\label{fig:drd-video}
	
	\vspace{-1.5em}
\end{figure*}

Taking bpp as the horizontal axis and the reconstructed \textit{PSNR, MS-SSIM, VMAF (MSE/MS-SSIM model)} as the vertical axis, we present the rate and distortion curves of different image compression schemes over USTC-TD 2022 and 2023 image datasets in Fig.~\ref{fig:img_results_avg}\,. From the overall results, we can find that the partially learned schemes (\textit{iWave++}\cite{ma2020end}, \textit{ELIC}\cite{he2022elic},  \textit{MLIC++}\cite{jiang2023mlic}) can outperform the traditional image compression schemes and achieve better compression performance than \textit{H.266/VVC} on proposed datasets under different metrics, which show its powerful potential. Here we analyze their performance from the perspective of content factors of different test images of USTC-TD. As shown in Fig.~\ref{fig:drd-image}\,, the detailed RD curves of some test images (\textit{USTC-2022-12}, \textit{USTC-2023-15}, \textit{USTC-2022-17}, \textit{USTC-2023-07}) with some special phenomenons are illustrated, and the results of each test image are presented in supplementary material. Based on the performance comparison of these schemes and feature analysis of proposed datasets (Section IV.C), the conclusions mainly include four aspects:

\begin{itemize}
\item (1) The learned schemes show the good potential on some complex scenarios. For example, as shown in the Fig.~\ref{fig:drd-image}\, (a) and (b), compared to the overall results (Fig.~\ref{fig:img_results_avg}\,), more learned schemes (\textit{Autoregressive Model}\cite{minnen2018joint}, \textit{Cheng2020}\cite{cheng2020learned}, \textit{iWave++}\cite{ma2020end}, \textit{ELIC}\cite{he2022elic},  \textit{MLIC++}\cite{jiang2023mlic}) perform better than the traditional schemes on some test images with specific features of environment-related factors (Table~\ref{tab:image_metrics}\,), such as the \textit{USTC-2022-12} with the lower scores of CF, the \textit{USTC-2023-15} with the higher scores of SI. Meanwhile, the results also demonstrate the previous image train/test datasets \cite{Kodak,asuni2014testimages,clic} can guide the researcher to train/evaluate the basic ability of intra-frame redundancy removal of their schemes to some extent. 

\item (2) The traditional schemes show the powerful generalization ability in some extreme scenarios. For example, as shown in the Fig.~\ref{fig:drd-image}\, (c) and (d), compared to the overall results (Fig.~\ref{fig:img_results_avg}\,), the generalization ability of learned schemes is lacking to handle the evaluative images with extreme mixture features of environment/imaging-related factors well (Table~\ref{tab:image_metrics}\,), such as the  \textit{USTC-2022-17} with the lower scores of SI, CF and the higher scores of LI, the \textit{USTC-2023-07} with the lower scores of SI and the higher scores of LI. Although the performance of learned schemes surpasses the traditional schemes in general, they are still limited to some extreme scenarios. 

\item (3) Based on the analysis of different features (SI, CF, LI, TI) of proposed datasets (Section IV.C), the detailed characteristics of different test images can efficiently assist the researcher to analyze the detailed bottleneck of their compression scheme.

\item (4) Compared to the performance of these schemes on different image datasets\cite{Kodak, asuni2014testimages, clic},  the proposed datasets can collaborate with other datasets to handle a wide coverage of performance evaluation, which also demonstrates the efficiency of the specific design of the proposed datasets' different content factors/features.\vspace{-0.4em}
\end{itemize}

\begin{table}
	\renewcommand\arraystretch{1.4}
	\centering
	\fontsize{6.7pt}{8pt}\selectfont
	\captionsetup{justification=centering}
	\caption{The Overall \textit{MOS} Results of Compressed Images of Classic Standardized and Advanced Learned Image Compression Schemes, where \textcolor{blue}{Blue} Represents the Lowest Score and \textcolor{red}{Red} Represents the Highest Score.}
	\label{tab:image_mos}
	\vspace{-0.5em}
	\begin{threeparttable}
		\setlength{\tabcolsep}{2.5mm}
		{
			\begin{tabular}{c|c|c|c}
				\hline
				\textbf{Dataset}                                                                                                              & \textbf{\begin{tabular}[c]{@{}c@{}}Scheme\\ Classification\end{tabular}} & \textbf{\begin{tabular}[c]{@{}c@{}}Compression\\ Scheme\end{tabular}} & \textbf{\begin{tabular}[c]{@{}c@{}}MOS\\ Score\end{tabular}} \\ \hline
				\multicolumn{1}{c|}{\multirow{6}{*}{\textbf{\begin{tabular}[c]{@{}c@{}}USTC-TD\\ 2022 \& 2023\\ Image Dataset\end{tabular}}}} & \multirow{1}{*}{Traditional}                                             & \textit{H.266/VVC} \cite{bross2021overview}                                                       &  3.67                                                            \\ \cline{2-4}
				\multicolumn{1}{c|}{}                                                                                                         &   \multirow{5}{*}{Learned}                                                 & \textit{Factorized Model}\cite{balle2018variational}                                             &                                                           3.44   \\ \cline{3-4} 
				\multicolumn{1}{c|}{}                                                                                                         &                                                                          & \textit{Hyperprior Model}\cite{balle2018variational}                                              &     \textcolor{blue}{3.43}                                                         \\ \cline{3-4} 
				\multicolumn{1}{c|}{}                                                                                                         &                                                                          & \textit{Autoregressive Model}\cite{minnen2018joint}                                         &    3.65                                                          \\ \cline{3-4} 
				\multicolumn{1}{c|}{}                                                                                                         &                                                                          & \textit{Cheng2020}\cite{cheng2020learned}                                                    &      3.77                                                        \\ \cline{3-4} 
				\multicolumn{1}{c|}{}                                                                                                         &                                                                          & \textit{MLIC++} \cite{jiang2023mlic}                                                      &    \textcolor{red}{3.80}                                                          \\ \hline
		\end{tabular}}
	\end{threeparttable}
	\vspace{-1.5em}
\end{table}

\subsubsection{Subjective Quality Evaluation and Analysis of Advanced Image Compression Schemes on USTC-TD}
In Table~\ref{tab:image_mos}\,, we present the overall \textit{MOS} results of compressed images of different image compression schemes over USTC-TD 2022 and 2023 image datasets. From the overall results, we can find that recent advanced learned compression schemes \cite{cheng2020learned, jiang2023mlic} outperform the traditional compression schemes, which demonstrates their powerful potential. This can be attributed to the optimization of \textit{MS-SSIM} for perceptual quality and the fact that end-to-end learning methods, compared to hand-crafted approaches, are more flexible and can be easily optimized toward different quality assessment metrics. From the detailed results of compressed images of different schemes mentioned in supplementary material, it is observed that different compression schemes show different \textit{MOS} performance trends on different images. Here we analyze their performance from the perspective of feature analysis of different test images of USTC-TD (Section IV.C), and explore the advantages of different learned compression schemes for varying scenes under the evaluation of subjective metric. The conclusions include the following two aspects:

\begin{itemize}	
	\item (1) The learned schemes show the powerful potential on various scenarios for perception optimization, especially the complex scenarios with mixture features, such as the \textit{USTC-2023-03} with the lower SI and higher CF scores, \textit{USTC-2023-05} with the higher SI, lower CF and LI scores. Combined with the performance of objective quality metrics, these samples highlight the learned schemes' generalization in optimizing across different metrics.
	
	\item (2) Despite traditional schemes\cite{bross2021overview} being hand-crafted and optimized for \textit{PSNR}, their subjective optimization capability remains competitive with certain learned schemes. Moreover, they exhibit a degree of practical applicability for visual perception, achieving comparable performance to some state-of-the-art learned schemes \cite{cheng2020learned, jiang2023mlic} in challenging scenarios, such as \textit{USTC-2022-02} with lower SI and LI scores, and \textit{USTC-2023-15} with higher SI and lower LI scores. Therefore, for the development of learned schemes, merely modifying the optimization metric is not sufficient; it is essential to further integrate perceptual theories into the framework design to achieve more effective improvements.
\end{itemize}

\begin{table*}
	\renewcommand\arraystretch{1.05}
	\centering
	\fontsize{6.7pt}{8pt}\selectfont
	\vspace{-2em}
	\caption{BD-rate (\%) Comparison for PSNR. Short Setting with Intra Period = 32. The anchor is VTM.}
	\vspace{-1em}
	\begin{threeparttable}
		\setlength{\tabcolsep}{1.7mm}
		{
			\begin{tabular}{cccccccccccccccccc}
				\hline
				\textbf{Dataset}       & \multicolumn{1}{l}{\textbf{\textit{VTM}}} & \textbf{\textit{HM}} &
				\textbf{\textit{AV1}} &
				\textbf{\textit{AV2}} & \multicolumn{1}{l}{\textbf{\textit{DVC\_Pro}}} & \multicolumn{1}{l}{\textbf{\textit{DCVC}}} & \multicolumn{1}{l}{\textbf{\textit{CANF-VC}}} & \textbf{\textit{TCM-VC}} & \textbf{\textit{VNVC}} & \textbf{\textit{DCVC-HEM}} & \textbf{\textit{OOFE}} & \textbf{\textit{SDD}} & \textbf{\textit{DCVC-DC}} & \textbf{\textit{DCVC-FM}}\\ \hline
				\multicolumn{1}{c|}{HEVC Class B}     & 0.0                              & 39.0      & -- & --                            & 188.6                                 & 115.7                             & 58.2                                 & 32.8       & 24.6     & -0.7    & -15.3  & -13.7 & -13.9  & -8.8       \\
				\multicolumn{1}{c|}{HEVC Class C}     & 0.0                              & 37.6    & -- & --                 & 202.8                                 & 150.8                             & 73.0                                 & 62.1      & 48.2      & 16.1      & -2.2  & -2.3   & -8.8  & -5.0   \\
				\multicolumn{1}{c|}{HEVC Class D}     & 0.0                              & 34.7     & -- & --                       & 160.3                                 & 106.4                             & 48.8                                 & 29.0        & 22.4    & -7.1        & -23.0   & -24.9 & -27.7 & -23.3   \\
				\multicolumn{1}{c|}{HEVC Class E}     & 0.0                              & 48.6    & -- & --              & 429.5                                 & 257.5                             & 116.8                                & 75.8     & 66.8       & 20.9        & -0.4   & -8.4 & -19.1 & -20.8 \\
				\multicolumn{1}{c|}{HEVC Class RGB}   & 0.0                              & 44.0   & -- & --                  & 186.8                                 & 118.6                             & 87.5                                 & 25.4       & 16.0     & -15.6        & -17.5   & -17.5 & -27.9 & -18.6 \\
				\multicolumn{1}{c|}{UVG}              & 0.0                              & 36.4       & -- & --                       & 218.7                                 & 129.5                             & 56.3                                 & 23.1      & 18.0      & -17.2      & -22.3    &  -19.7  & -25.9 & -20.5 \\
				\multicolumn{1}{c|}{MCL-JCV}          & 0.0                              & 41.9      & -- & --                        & 163.6                                 & 103.9                             & 60.5                                 & 38.2      & 30.2      & -1.6      & -5.8    &  -7.1   & -14.4 & -7.4 \\
				\multicolumn{1}{c|}{\textbf{\textcolor{blue}{USTC-TD\tnote{1}}}} & \textbf{\textcolor{blue}{0.0}}                              & \textbf{\textcolor{blue}{46.7}}                       
				& \textbf{\textcolor{blue}{34.3}}                        
				& \textbf{\textcolor{blue}{11.0}}                          & \textbf{\textcolor{blue}{284.4}}                            & \textbf{\textcolor{blue}{124.7}}                        & \textbf{\textcolor{blue}{64.6}}                           & \textbf{\textcolor{blue}{67.1}} & \textbf{\textcolor{blue}{60.4}}     & \textbf{\textcolor{blue}{16.0}}  & \textbf{\textcolor{blue}{8.2}}  & \textbf{\textcolor{blue}{3.7}} & \textbf{\textcolor{blue}{6.3}}  & \textbf{\textcolor{blue}{24.9}}    \\ \hline
				\multicolumn{1}{c|}{\textcolor{red}{\textbf{Average (\textit{Desirable})}}}          & \textbf{\textcolor{red}{0.0}}                              & \textbf{\textcolor{red}{41.8}}                       &
				\textbf{\textcolor{red}{--}}                       &
				\textbf{\textcolor{red}{--}}                       & \textbf{\textcolor{red}{230.9}}                            & \textbf{\textcolor{red}{139.7}}                        & \textbf{\textcolor{red}{72.8}}                           & \textbf{\textcolor{red}{47.2}}  & \textbf{\textcolor{red}{38.4}}    & \textbf{\textcolor{red}{4.0}}    & \textbf{\textcolor{red}{-8.0}} & \textbf{\textcolor{red}{-10.0}}& \textbf{\textcolor{red}{-15.1}} & \textbf{\textcolor{red}{-8.4}}   \\ \hline
				\multicolumn{1}{c|}{\textcolor{red}{\textbf{Average (\textit{Ideal})}}}          & \textbf{\textcolor{red}{0.0}}                              & \textbf{\textcolor{red}{41.1}}                       &
				\textbf{\textcolor{red}{--}}                       &
				\textbf{\textcolor{red}{--}}                       & \textbf{\textcolor{red}{229.3}}                            & \textbf{\textcolor{red}{138.4}}                        & \textbf{\textcolor{red}{70.7}}                           & \textbf{\textcolor{red}{44.2}}  & \textbf{\textcolor{red}{35.8}}    & \textbf{\textcolor{red}{1.4}}    & \textbf{\textcolor{red}{-9.8}} & \textbf{\textcolor{red}{-11.2}}& \textbf{\textcolor{red}{-16.4}} & \textbf{\textcolor{red}{-9.9}}   \\ \hline
			\end{tabular}
		}
	\begin{tablenotes}    
		\fontsize{6pt}{8pt}\selectfont
		\item[1] The results of USTC-TD indicate the average results of the \textit{Challenging} preset of dataset collaboration mentioned in Table~\ref{tab:dataset_preset}\,\,. 
	\end{tablenotes}
		\vspace{1em}
		\label{ip32_96_psnr}
	\end{threeparttable}
	
	\renewcommand\arraystretch{1.05}
	\centering
	\caption{BD-rate (\%) Comparison for MS-SSIM. Short Setting with Intra Period = 32. The anchor is VTM.}
	\vspace{-1em}
	\begin{threeparttable}
		\setlength{\tabcolsep}{1.7mm}
		{
			\begin{tabular}{cccccccccccccccccc}
				\hline
				\textbf{Dataset}                      & \multicolumn{1}{l}{\textbf{\textit{VTM}}} & \multicolumn{1}{c}{\textbf{\textit{HM}}}  &
				\multicolumn{1}{c}{\textbf{\textit{AV1}}}  &
				\multicolumn{1}{c}{\textbf{\textit{AV2}}}  & \multicolumn{1}{l}{\textbf{\textit{DVC\_Pro}}} & \multicolumn{1}{l}{\textbf{\textit{DCVC}}} & \multicolumn{1}{l}{\textbf{\textit{CANF-VC}}} & \multicolumn{1}{l}{\textbf{\textit{TCM-VC}}} & \multicolumn{1}{l}{\textbf{\textit{VNVC}}} & \multicolumn{1}{l}{\textbf{\textit{DCVC-HEM}}} & \multicolumn{1}{l}{\textbf{\textit{OOFE}}} & \multicolumn{1}{l}{\textbf{\textit{SDD}}} & \multicolumn{1}{l}{\textbf{\textit{DCVC-DC}}} & \multicolumn{1}{l}{\textbf{\textit{DCVC-FM}}}\\ \hline
				\multicolumn{1}{c|}{HEVC Class B}     & 0.0                              & 36.8         & -- & --                       & 67.0                                  & 35.9                              & 25.5                                 & -20.5                            & -33.1   & -47.4    &  -15.1   &  -48.0  &  -53.0   &  -12.5                     \\
				\multicolumn{1}{c|}{HEVC Class C}     & 0.0                              & 38.7     & -- & --                       & 61.1                                  & 24.9                              & 17.7                                 & -21.7         & -29.3                      & -43.3        & -15.9             & -49.6                            & -54.6                                    & -18.0                          \\
				\multicolumn{1}{c|}{HEVC Class D}     & 0.0                              & 34.9    & -- & --                         & 25.3                                  & 2.7                               & 1.5                                  & -36.2    & -41.1                           & -55.5          & -28.6             & -60.0                            & -63.4                                    & -30.6                                                    \\
				\multicolumn{1}{c|}{HEVC Class E}     & 0.0                              & 38.4        & -- & --                      & 195.8                                 & 90.0                              & 114.9                                & -20.5    & -0.4                           & -52.4       & -9.3             & -51.5                            & -60.7                                    & -32.6                                                          \\
				\multicolumn{1}{c|}{HEVC Class RGB}   & 0.0                              & 37.3         & -- & --                     & 66.8                                  & 43.7                              & 52.9                                 & -21.1    & -32.4                           & -45.8        & -16.7             & -46.3                            & -54.4                                    & -16.6                                               \\
				\multicolumn{1}{c|}{UVG}              & 0.0                              & 37.1         & -- & --                      & 74.6                                  & 11.9                              & 33.1                                 & -6.0   & -15.2                             & -32.7             & -10.6             & \multicolumn{1}{c}{-34.2}        & -36.7                                    & -7.3                                                     \\
				\multicolumn{1}{c|}{MCL-JCV}          & 0.0                              & 43.7        & -- & --                       & 46.1                                 & 39.1                              & 11.7                                 & -18.6   & -29.0                            & -44.0               & -2.5             & -46.3                            & -49.1                                    & -5.0                                                  \\
				\multicolumn{1}{c|}{\textbf{\textcolor{blue}{USTC-TD\tnote{1}}}} & \textbf{\textcolor{blue}{0.0}}                              & \textbf{\textcolor{blue}{47.7}}  
				& \textbf{\textcolor{blue}{31.8}} 
				& \textbf{\textcolor{blue}{11.0}} 
				& \textbf{\textcolor{blue}{55.8}}                                     & \textbf{\textcolor{blue}{3.9}}                                 & \textbf{\textcolor{blue}{3.4}}                                   & \textbf{\textcolor{blue}{-19.1}} & \textbf{\textcolor{blue}{-23.7}}                        & \textbf{\textcolor{blue}{-44.3}}      & \textbf{\textcolor{blue}{0.4}}    & \textbf{\textcolor{blue}{-48.0}}   & \textbf{\textcolor{blue}{-48.7}}                           & \textbf{\textcolor{blue}{39.5}}                          \\ \hline
				\multicolumn{1}{c|}{\textbf{\textcolor{red}{Average (\textit{Desirable})}}} & \textbf{\textcolor{red}{0.0}}                              & \textbf{\textcolor{red}{39.6}}                          & \textbf{\textcolor{red}{--}}                            & \textbf{\textcolor{red}{--}}                                     & \textbf{\textcolor{red}{74.0}}                                     & \textbf{\textcolor{red}{34.3}}                                 & \textbf{\textcolor{red}{32.5}}                                    & \textbf{\textcolor{red}{-22.5}}  & \textbf{\textcolor{red}{-27.1}}                          & \textbf{\textcolor{red}{-47.5}}       & \textbf{\textcolor{red}{-12.5}}     & \textbf{\textcolor{red}{-50.0}}    & \textbf{\textcolor{red}{-54.8}}                            & \textbf{\textcolor{red}{-10.8}}                           \\ \hline
				\multicolumn{1}{c|}{\textbf{\textcolor{red}{Average (\textit{Ideal})}}} & \textbf{\textcolor{red}{0.0}}                              & \textbf{\textcolor{red}{39.3}}                     & \textbf{\textcolor{red}{--}}          & \textbf{\textcolor{red}{--}}                        & \textbf{\textcolor{red}{74.1}}                                     & \textbf{\textcolor{red}{31.5}}                                 & \textbf{\textcolor{red}{32.6}}                                    & \textbf{\textcolor{red}{-20.5}}  & \textbf{\textcolor{red}{-25.5}}                          & \textbf{\textcolor{red}{-45.7}}      & \textbf{\textcolor{red}{-12.3}}     & \textbf{\textcolor{red}{-48.0}}    & \textbf{\textcolor{red}{-52.6}}                            & \textbf{\textcolor{red}{-10.4}}                           \\ \hline
			\end{tabular}
		}
		\label{ip32_96_ms-ssim}
	\begin{tablenotes}    
		\fontsize{6pt}{8pt}\selectfont
		\item[1] The results of USTC-TD indicate the average results of the \textit{Challenging} preset of dataset collaboration mentioned in Table~\ref{tab:dataset_preset}\,\,. 
	\end{tablenotes}
	\end{threeparttable}

	\vspace{1.2em}
	\renewcommand\arraystretch{1.2}
	\centering
	\caption{BD-rate (\%) Comparison for VMAF (MSE model), VMAF (MS-SSIM model). \\Short Setting with Intra Period = 32. The anchor is VTM.}
	\vspace{-1em}
	\begin{threeparttable}
		\setlength{\tabcolsep}{1.1mm}
		{
			\begin{tabular}{c|c|clclclclclclclclclclclclclcl}
				\hline
				\textbf{Dataset}                  & \textbf{Evaluative Metric} & \multicolumn{2}{c}{\textit{\textbf{VTM}}} & \multicolumn{2}{c}{\textit{\textbf{HM}}} &
				\multicolumn{2}{c}{\textit{\textbf{AV1}}} &
				\multicolumn{2}{c}{\textit{\textbf{AV2}}} & \multicolumn{2}{c}{\textit{\textbf{DVC\_Pro}}} & \multicolumn{2}{c}{\textit{\textbf{DCVC}}} & \multicolumn{2}{c}{\textit{\textbf{CANF-VC}}} & \multicolumn{2}{c}{\textit{\textbf{TCM-VC}}} & \multicolumn{2}{c}{\textit{\textbf{VNVC}}} & \multicolumn{2}{c}{\textit{\textbf{DCVC-HEM}}} & \multicolumn{2}{c}{\textit{\textbf{OOFE}}} & \multicolumn{2}{c}{\textit{\textbf{SDD}}} & \multicolumn{2}{c}{\textit{\textbf{DCVC-DC}}} & \multicolumn{2}{c}{\textit{\textbf{DCVC-FM}}} \\ \hline
				\multirow{2}{*}{\textbf{USTC-TD}} & VMAF (MSE model)           & \multicolumn{2}{c}{\textbf{0.0}}          & \multicolumn{2}{c}{\textbf{46.5}}         &
				\multicolumn{2}{c}{\textbf{291.9}}         &
				\multicolumn{2}{c}{\textbf{137.1}}         & \multicolumn{2}{c}{\textbf{118.5}}               & \multicolumn{2}{c}{\textbf{63.0}}           & \multicolumn{2}{c}{\textbf{39.6}}              & \multicolumn{2}{c}{\textbf{45.8}}             & \multicolumn{2}{c}{\textbf{42.2}}           & \multicolumn{2}{c}{\textbf{10.6}}               & \multicolumn{2}{c}{\textbf{9.5}}           & \multicolumn{2}{c}{\textbf{-1.1}}          & \multicolumn{2}{c}{\textbf{9.2}}              & \multicolumn{2}{c}{\textbf{21.6}}              \\ \cline{2-30} 
				& VMAF (MS-SSIM model)       & \multicolumn{2}{c}{\textbf{--}}          & \multicolumn{2}{c}{\textbf{--}}         &
				\multicolumn{2}{c}{\textbf{--}}         &
				\multicolumn{2}{c}{\textbf{--}}         & \multicolumn{2}{c}{\textbf{160.6}}               & \multicolumn{2}{c}{\textbf{89.4}}           & \multicolumn{2}{c}{\textbf{93.7}}              & \multicolumn{2}{c}{\textbf{107.3}}             & \multicolumn{2}{c}{\textbf{112.8}}           & \multicolumn{2}{c}{\textbf{64.8}}               & \multicolumn{2}{c}{\textbf{--}}           & \multicolumn{2}{c}{\textbf{46.1}}          & \multicolumn{2}{c}{\textbf{34.5}}              & \multicolumn{2}{c}{\textbf{--}}              \\ \hline
			\end{tabular}
			\vspace{1.2em}
		}
		\label{ip32_96_vmaf}
	\end{threeparttable}
	
	\renewcommand\arraystretch{1.2}
	\centering
	\caption{BD-rate (\%) Comparison for PSNR, MS-SSIM, VMAF (MSE model), VMAF (MS-SSIM model). \\Long Setting with Intra Period = 32. The anchor is VTM.}
	\vspace{-0.8em}
	\begin{threeparttable}
		\setlength{\tabcolsep}{1.6mm}
		{
			\begin{tabular}{c|c|clclclclclclclclclclclcl}
				\hline
				\textbf{Dataset}                  & \textbf{Evaluative Metric} & \multicolumn{2}{c}{\textit{\textbf{VTM}}} & 
				\multicolumn{2}{c}{\textit{\textbf{AV1}}} &
				\multicolumn{2}{c}{\textit{\textbf{AV2}}} & \multicolumn{2}{c}{\textit{\textbf{DVC\_Pro}}} & \multicolumn{2}{c}{\textit{\textbf{DCVC}}} & \multicolumn{2}{c}{\textit{\textbf{CANF-VC}}} & \multicolumn{2}{c}{\textit{\textbf{TCM-VC}}} &  \multicolumn{2}{c}{\textit{\textbf{DCVC-HEM}}} & \multicolumn{2}{c}{\textit{\textbf{OOFE}}} & \multicolumn{2}{c}{\textit{\textbf{SDD}}} & \multicolumn{2}{c}{\textit{\textbf{DCVC-DC}}} & \multicolumn{2}{c}{\textit{\textbf{DCVC-FM}}} \\ \hline
				\multirow{4}{*}{\textbf{USTC-TD}} & PSNR                       & \multicolumn{2}{c}{\textbf{0.0}}          & 
				\multicolumn{2}{c}{\textbf{33.6}}         &
				\multicolumn{2}{c}{\textbf{11.0}}         & \multicolumn{2}{c}{\textbf{403.5}}               & \multicolumn{2}{c}{\textbf{120.8}}           & \multicolumn{2}{c}{\textbf{70.2}}              & \multicolumn{2}{c}{\textbf{54.4}}             &  \multicolumn{2}{c}{\textbf{9.5}}               & \multicolumn{2}{c}{\textbf{2.8}}           & \multicolumn{2}{c}{\textbf{-1.5}}          & \multicolumn{2}{c}{\textbf{0.4}}              & \multicolumn{2}{c}{\textbf{13.5}}              \\ \cline{2-26} 
				& MS-SSIM                    & \multicolumn{2}{c}{\textbf{0.0}}          & 
				\multicolumn{2}{c}{\textbf{28.4}}         &
				\multicolumn{2}{c}{\textbf{7.0}}         & \multicolumn{2}{c}{\textbf{43.3}}               & \multicolumn{2}{c}{\textbf{5.6}}           & \multicolumn{2}{c}{\textbf{5.7}}              & \multicolumn{2}{c}{\textbf{-25.3}}                     & \multicolumn{2}{c}{\textbf{-49.1}}               & \multicolumn{2}{c}{\textbf{-3.0}}           & \multicolumn{2}{c}{\textbf{-51.5}}          & \multicolumn{2}{c}{\textbf{-53.4}}              & \multicolumn{2}{c}{\textbf{23.5}}              \\ \cline{2-26} 
				& VMAF (MSE model)           & \multicolumn{2}{c}{\textbf{0.0}}          & 
				\multicolumn{2}{c}{\textbf{349.2}}         &
				\multicolumn{2}{c}{\textbf{194.5}}         & \multicolumn{2}{c}{\textbf{118.9}}               & \multicolumn{2}{c}{\textbf{58.0}}           & \multicolumn{2}{c}{\textbf{23.8}}              & \multicolumn{2}{c}{\textbf{33.7}}             &  \multicolumn{2}{c}{\textbf{4.5}}               & \multicolumn{2}{c}{\textbf{3.5}}           & \multicolumn{2}{c}{\textbf{-6.7}}          & \multicolumn{2}{c}{\textbf{2.4}}              & \multicolumn{2}{c}{\textbf{10.5}}              \\ \cline{2-26} 
				& VMAF (MS-SSIM model)       & \multicolumn{2}{c}{\textbf{--}}          & 
				\multicolumn{2}{c}{\textbf{--}}         &
				\multicolumn{2}{c}{\textbf{--}}         & \multicolumn{2}{c}{\textbf{166.8}}               & \multicolumn{2}{c}{\textbf{104.6}}           & \multicolumn{2}{c}{\textbf{94.8}}              & \multicolumn{2}{c}{\textbf{111.0}}             &  \multicolumn{2}{c}{\textbf{69.9}}               & \multicolumn{2}{c}{\textbf{--}}           & \multicolumn{2}{c}{\textbf{52.1}}          & \multicolumn{2}{c}{\textbf{37.8}}              & \multicolumn{2}{c}{\textbf{--}}              \\ \hline
			\end{tabular}
			\vspace{1.2em}
		}
		\label{ip32_300}
	\end{threeparttable}
	
	\renewcommand\arraystretch{1.2}
	\centering
	\caption{BD-rate (\%) Comparison for PSNR, MS-SSIM, VMAF (MSE model), VMAF (MS-SSIM model). \\Long Setting with Intra Period = -1. The anchor is VTM.}
	\vspace{-0.8em}
	\begin{threeparttable}
		\setlength{\tabcolsep}{1.6mm}
		{
		\begin{tabular}{c|c|clclclclclclclclclclclcl}
			\hline
			\textbf{Dataset}                  & \textbf{Evaluative Metric} & \multicolumn{2}{c}{\textit{\textbf{VTM}}} & 
			\multicolumn{2}{c}{\textit{\textbf{AV1}}} &
			\multicolumn{2}{c}{\textit{\textbf{AV2}}} & \multicolumn{2}{c}{\textit{\textbf{DVC\_Pro}}} & \multicolumn{2}{c}{\textit{\textbf{DCVC}}} & \multicolumn{2}{c}{\textit{\textbf{CANF-VC}}} & \multicolumn{2}{c}{\textit{\textbf{TCM-VC}}} &  \multicolumn{2}{c}{\textit{\textbf{DCVC-HEM}}} & \multicolumn{2}{c}{\textit{\textbf{OOFE}}} & \multicolumn{2}{c}{\textit{\textbf{SDD}}} & \multicolumn{2}{c}{\textit{\textbf{DCVC-DC}}} & \multicolumn{2}{c}{\textit{\textbf{DCVC-FM}}} \\ \hline
			\multirow{4}{*}{\textbf{USTC-TD}} & PSNR                       & \multicolumn{2}{c}{\textbf{0.0}}          & 
			\multicolumn{2}{c}{\textbf{33.4}}         &
			\multicolumn{2}{c}{\textbf{-1.9}}         & \multicolumn{2}{c}{\textbf{520.7}}               & \multicolumn{2}{c}{\textbf{221.4}}           & \multicolumn{2}{c}{\textbf{95.6}}              & \multicolumn{2}{c}{\textbf{99.1}}             &  \multicolumn{2}{c}{\textbf{23.9}}               & \multicolumn{2}{c}{\textbf{16.5}}           & \multicolumn{2}{c}{\textbf{15.2}}          & \multicolumn{2}{c}{\textbf{9.2}}              & \multicolumn{2}{c}{\textbf{22.4}}              \\ \cline{2-26} 
			& MS-SSIM                    & \multicolumn{2}{c}{\textbf{0.0}}          & 
			\multicolumn{2}{c}{\textbf{24.4}}         &
			\multicolumn{2}{c}{\textbf{-5.0}}         & \multicolumn{2}{c}{\textbf{64.9}}               & \multicolumn{2}{c}{\textbf{30.7}}           & \multicolumn{2}{c}{\textbf{17.5}}              & \multicolumn{2}{c}{\textbf{-2.3}}             &  \multicolumn{2}{c}{\textbf{-41.3}}               & \multicolumn{2}{c}{\textbf{10.7}}           & \multicolumn{2}{c}{\textbf{-42.5}}          & \multicolumn{2}{c}{\textbf{-51.2}}              & \multicolumn{2}{c}{\textbf{34.7}}              \\ \cline{2-26} 
			& VMAF (MSE model)           & \multicolumn{2}{c}{\textbf{0.0}}          & 
			\multicolumn{2}{c}{\textbf{387.8}}         &
			\multicolumn{2}{c}{\textbf{95.7}}         & \multicolumn{2}{c}{\textbf{195.4}}               & \multicolumn{2}{c}{\textbf{100.0}}           & \multicolumn{2}{c}{\textbf{49.1}}              & \multicolumn{2}{c}{\textbf{59.2}}             &  \multicolumn{2}{c}{\textbf{17.3}}               & \multicolumn{2}{c}{\textbf{15.1}}           & \multicolumn{2}{c}{\textbf{4.1}}          & \multicolumn{2}{c}{\textbf{10.4}}              & \multicolumn{2}{c}{\textbf{18.7}}              \\ \cline{2-26} 
			& VMAF (MS-SSIM model)       & \multicolumn{2}{c}{\textbf{--}}          & 
			\multicolumn{2}{c}{\textbf{--}}         &
			\multicolumn{2}{c}{\textbf{--}}         & \multicolumn{2}{c}{\textbf{277.2}}               & \multicolumn{2}{c}{\textbf{132.7}}           & \multicolumn{2}{c}{\textbf{146.3}}              & \multicolumn{2}{c}{\textbf{158.0}}             &  \multicolumn{2}{c}{\textbf{99.0}}               & \multicolumn{2}{c}{\textbf{--}}           & \multicolumn{2}{c}{\textbf{82.9}}          & \multicolumn{2}{c}{\textbf{50.8}}              & \multicolumn{2}{c}{\textbf{--}}              \\ \hline
		\end{tabular}
			\vspace{-1.1em}
		}
		\label{ip-1_300}
	\end{threeparttable}
\end{table*}

\subsubsection{Objective Quality Evaluation and Analysis of Advanced Video Compression Schemes on USTC-TD}
Taking bpp as the horizontal axis and the reconstructed \textit{PSNR, MS-SSIM, VMAF (MSE/MS-SSIM model)} as the vertical axis, we present the rate and distortion curves and BD-rate results of different video compression schemes over the USTC-TD 2023 video dataset in Fig.~\ref{fig:video_results_avg}\,, Table~\ref{ip32_96_psnr}, \ref{ip32_96_ms-ssim}\,, \ref{ip32_96_vmaf}\, for short dataset, Table~\ref{ip32_300}\,, Table~\ref{ip-1_300}\, for long dataset. Note that the setting of intra period = -1 is used to enhance comprehensive benchmark for the testing of long sequences, it allows for an accurate evaluation of the video compression scheme’s robustness to handle the long video sequences. From the overall results of \textit{PSNR} and \textit{VMAF (MSE model)} metric, we can find that the performance of the advanced traditional schemes is better than that of all advanced learned schemes on the proposed short/long video dataset. Notably, the \textit{PSNR} results of short setting differ from the conclusions drawn from other datasets \cite{mercat2020uvg, wang2016mcl, bossen2013common, boyce2018jvet}. From the overall results of \textit{MS-SSIM} metric, the performance of the traditional schemes is lower than that of advanced learned schemes. Here we analyze their performance from the perspective of typical characteristics of different test video contents and test length of USTC-TD video dataset. As shown in Fig.~\ref{fig:drd-video}\,, the detailed RD curves of some test videos (\textit{USTC-FourPeople}, \textit{USTC-BasketballPass}, \textit{USTC-Snooker}, \textit{USTC-BicycleDriving}) with some special phenomenons are illustrated, the results of each test video are also presented in supplementary material. Based on the performance comparison of advanced schemes and feature analysis of proposed datasets (Section IV.C), the conclusions mainly include the following four aspects:\vspace{-0.1em}

\begin{itemize}
	\item (1) The traditional schemes show good generalization ability on various real-world video scenarios. As shown in Table~\ref{ip32_96_psnr}\,, different from the performance of advanced compression schemes on other datasets \cite{mercat2020uvg, wang2016mcl, bossen2013common, boyce2018jvet}\,, the traditional schemes still achieve the state-of-the-art performance of these schemes on USTC-TD under the PSNR metric. Different from the other datasets, USTC-TD video dataset focuses on various temporal correlation types. Combined the attributes of different test videos of USTC-TD (Fig.~\ref{fig:video_SI_TI}\,, Table~\ref{tab:VD2023}\,), we find that the traditional schemes can handle more scenarios with the various temporal features of motion-related elements (motion type, lens motion), such as the \textit{USTC-BicycleDriving} with higher scores of TI and \textit{USTC-Snooker} with the specific design of lens motion (Section IV.C). The performance of these test videos is shown in Fig.~\ref{fig:drd-video}\, (c) and (d), and the results demonstrate that the traditional schemes can handle these scenes with complex motion types robustly.  
	
	\item (2) The learned schemes show the optimistic potential on some scenarios with complex motion cases, such as the \textit{USTC-FourPeople} with lower TI scores, and the \textit{USTC-BasketballPass} with higher TI scores. These scenarios commonly  appear in the previous test datasets\cite{mercat2020uvg, wang2016mcl, bossen2013common, boyce2018jvet}\,, such as the \textit{FourPeople} and \textit{BasketballPass} in \textit{HEVC/VVC CTC}, which can guide the design and the optimized target of deep network to handle these motion situations. The results of these specific scenarios demonstrate the basic evaluative ability of the previous datasets and the performance potential of the deep learning-based manner. 
	
	\item (3) Based on the performance comparison of advanced compression schemes under different test length settings and intra period settings of USTC-TD (Table~\ref{ip32_96_psnr} for short dataset, Table~\ref{ip32_300}\, and Table~\ref{ip-1_300}\, for long dataset), the performance gap between traditional and learned schemes is amplified in the long setting, it further demonstrates the superiority of traditional schemes. Combined with the attributes of different length settings (Fig.~\ref{fig:video_SI_TI}\,), the more powerful temporal diversity of long sequences challenges the robustness of these schemes even further. The extended temporal variation with different intra period tests further exposes the shortcomings of advanced compression schemes, revealing the limitations of learned schemes in the error propagation of prediction chain, motion modeling, and the design of reference mechanism. 
	
	\item (4) Based on the analysis of video-related features (Section IV.C), our proposed dataset can make up more typical motion/temporal correlation-related real-world factors with other datasets to handle an excellent coverage of performance evaluation, which also demonstrates the efficiency of the specific design of the proposed datasets' different factors/features. Furthermore, compared to the performance of other datasets\cite{mercat2020uvg, wang2016mcl, bossen2013common, boyce2018jvet} of different schemes mentioned in \cite{lu2019dvc,ho2022canf, li2021deep, sheng2022temporal,li2022hybrid, tang2024offline,sheng2024spatial,sheng2024vnvc,li2023neural,li2024neural}, as shown in Table~\ref{ip32_96_psnr}\, and Table~\ref{ip32_96_ms-ssim}\,, the different phenomenon between the performance of our proposed dataset and other datasets also verifies the efficiency of the proposed dataset.
	
	\item (5) Based on the performance comparison of different presets of dataset collaborations under different feature coverage settings, as shown in Table~\ref{ip32_96_psnr} and \ref{ip32_96_ms-ssim}\,, the performance trends observed across different presets highlight their distinct testing preferences, and further validate the configuration rationality of different presets. Combining the attributes of each preset, the results from the \textit{Challenging} preset to the other presets gradually validate the performance of these video compression schemes, highlighting the testing effects with the consideration of general, reliable, and comprehensive evaluation, respectively. Specifically, in Table~\ref{ip32_96_psnr} and \ref{ip32_96_ms-ssim}\,, the performance of \textit{DVC\_Pro} and \textit{CANF-VC} tend to show variational BD-rate changes under different presets, reflecting their limitations in handling specific considerations or coverage settings compared to other methods. On the other hand, the variants of \textit{DCVC} show more consistent performance across different feature coverage settings, demonstrating that these methods are more robust when adapting to various scenarios. 		
\end{itemize}

\subsubsection{Subjective Quality Evaluation and Analysis of Advanced Video Compression Schemes on USTC-TD}
In Table~\ref{tab:video-mos}\,, we present the overall \textit{MOS} results of compressed videos of different video compression schemes over USTC-TD 2023 video dataset. The results of video compression follow the same trend as image compression, where learned schemes outperform traditional schemes, highlighting their advantages in perceptual quality optimization and adaptability to different quality assessment metrics. From the detailed results mentioned in the supplemental material, we present the detailed \textit{MOS} results of compressed videos of different video compression schemes over USTC-TD 2023 video dataset. It is also observed that different schemes show different \textit{MOS} performance trends on different videos. Here we further analyze their performance from the perspective of feature analysis of different test videos of USTC-TD (Section IV.C), and explore the advantages of different learned compression schemes for varying scenes under the evaluation of subjective metric. The conclusions mainly include the following two aspects:

\begin{itemize}	
	\item (1) The learned video schemes also demonstrate powerful potential for perceptual optimization, particularly in complex motion scenarios such as \textit{USTC-ShakingHands} and \textit{USTC-BasketballPass}, which exhibit higher TI scores. Combined with the performance of objective quality metrics, these samples further highlight the learned schemes' ability to generalize and optimize across different metrics.
	
	\item (2) Despite traditional schemes\cite{bross2021overview} being hand-crafted and optimized for \textit{PSNR}, they exhibit strong competitiveness compared to learned schemes. Especially in certain sequences with extreme motions, they demonstrate a greater advantage over learned schemes, such as \textit{USTC-BicycleDriving} with the highest TI scores, and \textit{USTC-BasketballDrill} with higher SI and TI scores. Therefore, for the development of learned schemes, merely modifying the optimization metric is also not enough. It is crucial to further incorporate video-related factors into the framework design, such as temporal consistency and motion fidelity, to achieve more effective improvements.
\end{itemize}

\vspace{-0.9em}
\subsection{Limitation and Inspiration of Advanced Image and Video Compression Schemes}
\vspace{-0.2em}
In this subsection, based on the analysis of experimental results, we  analyze the limitations of evaluative image/video compression schemes, and point out some limitations and inspirations among these advanced compression schemes. 

\begin{table}
	\renewcommand\arraystretch{1.35}
	\centering
	\fontsize{6.7pt}{8pt}\selectfont
	\captionsetup{justification=centering}
	\caption{The Overall \textit{MOS} Results of Compressed Videos of Classic Standardized and Advanced Learned Video Compression Schemes, where \textcolor{blue}{Blue} Represents the Lowest Score and \textcolor{red}{Red} Represents the Highest Score.}
	\vspace{-0.7em}
	\begin{threeparttable}
		\setlength{\tabcolsep}{2.5mm}
		{
			\begin{tabular}{c|c|c|c}
				\hline
				\textbf{Dataset}                                                                                  & \textbf{\begin{tabular}[c]{@{}c@{}}Scheme\\ Classification\end{tabular}} & \textbf{\begin{tabular}[c]{@{}c@{}}Compression\\ Scheme\end{tabular}} & \textbf{\begin{tabular}[c]{@{}c@{}}MOS\\ Score\end{tabular}} \\ \hline
				\multirow{8}{*}{\textbf{\begin{tabular}[c]{@{}c@{}}USTC-TD\\ 2023\\ Video Dataset\end{tabular}}} & \multirow{1}{*}{Traditional}                                                                                                                    & \textit{H.266/VVC}\cite{bross2021overview}                                                     &   3.49                                                            \\ \cline{2-4} 
				& \multirow{7}{*}{Learned}                                                 & 
				\textit{DCVC}\cite{li2021deep}                                                    &   \textcolor{blue}{3.14}                                                           \\ \cline{3-4} 
				\cline{3-4} 
				&                                                                          & \textit{TCM-VC}\cite{sheng2022temporal}                                                      &  3.24                                                            \\ \cline{3-4} 
				&                                                                          & \textit{DCVC-HEM}\cite{li2022hybrid}                                                     &   3.42                                                           \\ \cline{3-4} 
				&                                                                          & \textit{OOFE}\cite{tang2024offline}                                                        &  3.51                                                            \\ \cline{3-4} 
				&                                                                          & \textit{SDD}\cite{sheng2024spatial}                                                          &   3.41                                                           \\ \cline{3-4} 
				&                                                                          & \textit{DCVC-DC}\cite{li2023neural}                 & 3.51                                         \\ \cline{3-4} 
				&                                                                          & \textit{DCVC-FM}\cite{li2024neural}               & \textcolor{red}{3.55}                                         \\ \hline
			\end{tabular}
		}
	\end{threeparttable}
	\label{tab:video-mos}
	\vspace{-2em}
\end{table}

\begin{figure*}
	\centering
	\vspace{-0.6em}
	\includegraphics[width=145mm]{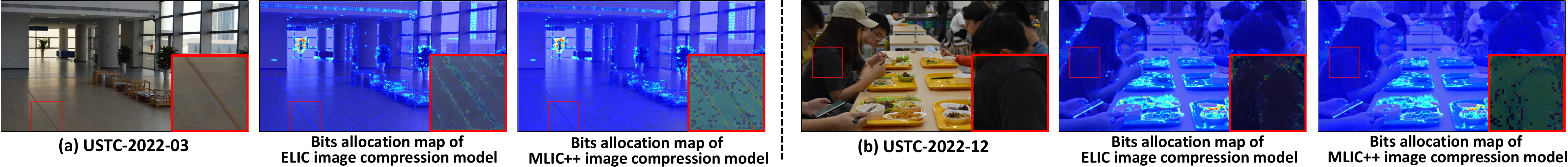}
	\vspace{-0.8em}
	\caption{Visualization of limitation of partial advanced image compression schemes (\textit{ELIC}\cite{he2022elic},  \textit{MLIC++}\cite{jiang2023mlic}) on \textit{USTC-2022-03} and \textit{USTC-2022-12}.}
	\label{fig:limitation_image}
\end{figure*}

\begin{figure*}
	\centering
	\vspace{-0.5em}
	\includegraphics[width=145mm]{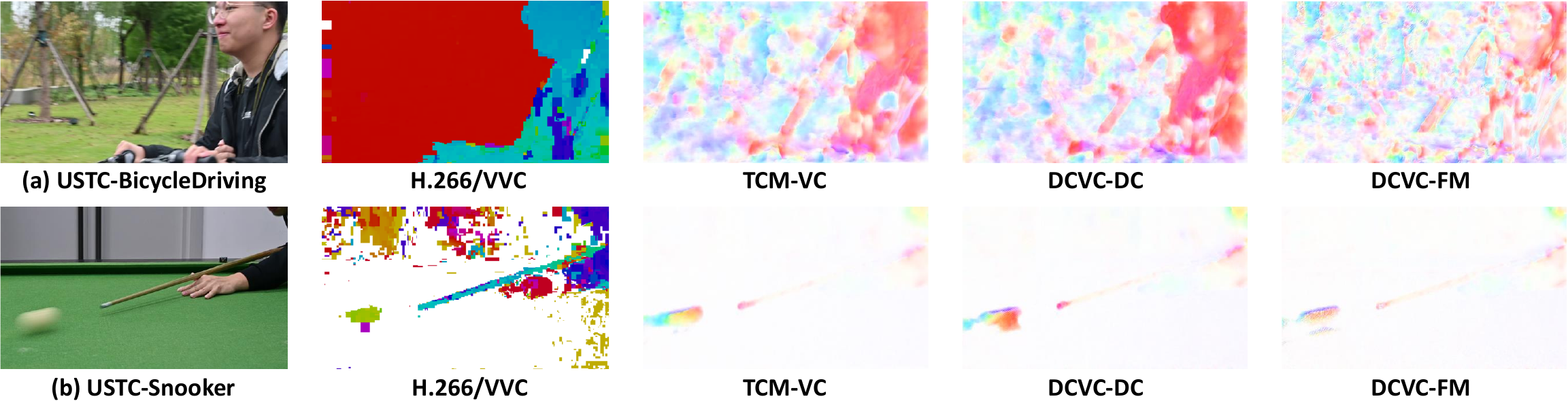}
	\vspace{-0.5em}
	\caption{Visualization of limitation of partial video compression schemes (\textit{TCM-VC}\cite{sheng2022temporal}, \textit{DCVC-DC}\cite{li2023neural}, \textit{DCVC-FM}\cite{li2024neural}) on \textit{USTC-BicycleDriving/Snooker}.}
	\label{limitation_video}
	\vspace{-1.5em}
\end{figure*}

\subsubsection{Limitation and Inspiration of Image Compression} 
Based on the above analysis in Section V.B, we can find that the generalization ability of the learned codec is a major challenge for practical usage. Most existing methods focus on improving the compression performance while neglecting its generalization for various scenarios. As the situations mentioned in the item (1) and (2) of conclusions (Section V.B (1)), the generalization ability is challenged with the effective extension of the evaluation data. These problems mainly arise from the incompletion of training data and the constraint optimization direction of deep learning-based manner with the limited evaluation data. Here we explore these problems based on the coding process of different compression schemes.

\begin{table}
	\renewcommand\arraystretch{1.3}
	\centering
	\scriptsize
	\caption{BD-rate (\%) Results of Different Training Strategies of Optical Flow-related Module of DCVC-DC \cite{li2023neural} \\ on USTC-TD under PSNR Metric}
	\vspace{-0.6em}
	\label{OOFE}
	\setlength{\tabcolsep}{6mm}
	{
		\begin{tabular}{ccl}
			\hline
			\multicolumn{1}{c|}{\textbf{Scheme}}        & \multicolumn{2}{c}{\textbf{BD-rate (\%)}}                                                        \\ \hline
			\multicolumn{1}{c|}{\textit{DCVC-DC}}        & \multicolumn{2}{c}{6.3\%}       \\ \hline
			\multicolumn{1}{c|}{\textit{DCVC-DC} + \textit{Flow Pre-training}}        & \multicolumn{2}{c}{-1.2\%} \\ \hline
		\end{tabular}
	}
	\vspace{-1.8em}
\end{table}

For example, as shown in Fig.~\ref{fig:img_results_avg}, the performance of one bpp point of \textit{MLIC++} is lower than that of  other schemes, but the performance of other bpp points is better. In detail, we further illustrate the detailed rate-distortion curves of each image, such as the RD curves of \textit{USTC-2022-12}, \textit{USTC-2022-17}, \textit{USTC-2023-07} shown in Fig.~\ref{fig:drd-image} (a), (b), and (c), we can find that one bpp model of \textit{MLIC++} all performs poorly on several specific images. To explore it, we visualize the bits allocation map of \textit{ELIC} and \textit{MLIC++} on some test images, such as the cases of \textit{USTC-2022-03} and \textit{USTC-2022-12} shown in Fig.~\ref{fig:limitation_image}\,. Compared to \textit{ELIC}, \textit{MLIC++} allocates more bits to some flat areas, whereas these areas could be encoded with fewer bits. Inspired by them, the controllable model optimization, domain adaptation, and precise rate allocation of the learned compression models need to be further improved for future practical usage.

\subsubsection{Limitation and Inspiration of Video Compression} 
Based on the above analysis in Section V.B, as shown in Fig.~\ref{fig:video_results_avg} and Table~\ref{ip32_96_psnr}\,, the performance of all learned video codecs is lower than that of traditional video codecs on USTC-TD, which is different from the performance phenomenon on other datasets. The reason mainly comes from that the learned video codecs perform poorly on some sequences with complex motion features. Here we further explore these problems based on the motion-related modules of different schemes.

As mentioned in Fig.~\ref{fig:drd-video}\, (c) and (d), the performance of the state-of-the-art learned video codecs is even lower than that of the \textit{H.265/HEVC}\cite{sullivan2012overview}. Based on these observations, we visualize the video reconstructed frames of these video codecs as shown in Fig.~\ref{limitation_video}\,. From the comparison of different scenarios, we can find that the specific design of motion-related features (high-speed moving objects, object occlusion, and camera motion) bring severe motion blur in the temporal domain, which challenges the optical flow-based motion estimation/compensation module of learned video codecs that is difficult to estimate accurate motion vector prediction. Therefore, we further illustrate the estimated motion vectors of these traditional and learned codecs in Fig.~\ref{limitation_video}\,, it obviously observes that the motion field of learned video compression schemes performs wrong and disordered, which further demonstrates that the flow-based motion-related modules of learned video codecs are difficult to handle the complex motion situations. 

To further verify the performance impact of these problems, we tentatively design the experiment to optimize the optical flow-related module of different learned video codecs. We set the state-of-the-art scheme (\textit{DCVC-DC} \cite{li2023neural}) as the anchor, and use the motion vectors of \textit{H.266/VVC} as the optimized target of the optical flow-based motion estimation module (\textit{Spynet}\cite{ranjan2017optical}) in the offline pre-training stage, instead of the usage of EPE loss for the training of these optical flow-based modules. The performance is shown in Table~\ref{OOFE}\,. Inspired by the results, it verifies that the motion modeling and training strategy of  learned video compression models are necessary to be further improved for practical usage in the future.

\vspace{-0.6em}
\section{Limitation Discussion}
\vspace{-0.2em}
Beyond the existing image/video datasets and USTC-TD, here we further discuss the limitations of these image/video datasets. With the collaboration of these datasets, the overall coverage of different image/video features has become more comprehensive, yet there are still some missing data types. 

First, one notable missing type is computer-generated (CG) content, such as animations, cartoon content, and screen content. Unlike the camera-captured content of existing datasets, these special images/sequences differ significantly in visual characteristics and elements, such as vibrant color schemes, exaggerated motion, and mixed textures. For example, as shown in Fig.~\ref{fig:video_SI_TI}\,, the particular sample of the existing datasets with the highest temporal diversity is the animated video of \textit{MCL-JCV} \cite{wang2016mcl}, which is the only sample from other datasets, aside from USTC-TD, with TI scores higher than 40. Second, AI-generated content \cite{duan2024pku, singer2022make} is
another crucial missing data type, which is created using advanced generative models, such as GANS and diffusion models. These images/videos introduce unique challenges for compression models, as they often contain novel visual patterns, surreal features, or unusual motion patterns, which significantly differ from the content captured by human creators. The absence of these above non-camera-captured content types reveals a limitation of existing datasets, which mainly focus on natural images/videos captured by cameras. As synthetic media becomes more common in fields like entertainment, gaming, digital art, and AI-generated content, the demand for their compression also increases, creating a need for datasets that include CG and AI-generated media. Therefore, addressing these limitations in the future would enhance compression models' ability to handle a broader range of visual content.

\vspace{-0.7em}
\section{Conclusion}
\vspace{-0.2em}
In this paper, we propose a test dataset (named USTC-TD) for compression-related research, which covers more diverse content factors. To evaluate the efficiency of USTC-TD, we qualitatively evaluate the USTC-TD on different image/video features and compare it with the previous image/video common test datasets to verify its excellent compensation. In addition, we evaluate the advanced compression schemes under different metrics benchmarked on USTC-TD, and further analyze their performance to point out the inspirations for future compression-related research.

In the present dataset construction process, we only consider the basic image and video test datasets. In the future, we plan to progressively extend the annotation datasets of USTC-TD for image/video coding for machine (ICM/VCM)\cite{duan2020video, gao2021towards, yan2021sssic}, such as object segmentation \cite{yang2024scalable}, object detection \cite{xu2023integrating}, action recognition \cite{feichtenhofer2017spatiotemporal}, \textit{et al}, and the reconstruction dataset of video enhancement, such as image/video super-resolution\cite{li2022mulut, xiao2021space}, denoising\cite{zhang2017beyond}, \textit{et al}, for the testing of compression-related downstream researches, and provide a comprehensive baseline to promote the development of compression-related diverse tasks. 

\vspace{-0.5em}
\bibliographystyle{IEEEtran}
\bibliography{IEEEexample}
\end{document}